\documentclass[]{article}

\usepackage{alanstyle}
\usepackage{graphicx}

\title{Exact Solutions for a GBM-type Stochastic Volatility Model having
	a Stationary Distribution}
\author{Alan L. Lewis\footnote{Newport Beach, California, USA; email: alewis@financepress.com}}

\begin{document}

\maketitle

\begin{abstract}
	We find various exact solutions for a new stochastic volatility (SV) model: the
	transition probability density, European-style option values, and (when it
	exists) the martingale defect. This may represent the first example of 
	 an SV model combining exact solutions, GBM-type volatility noise,
	 and a stationary volatility density.

\end{abstract}

\section{Introduction}
  We develop the following stochastic volatility (SV) model for the real-world (P-measure) evolution of a financial asset $S_t$, such as a broad-based stock index:
  
  \bea  \mbox{Under P}: \quad \left \{ 
  \begin{array}{ll}
  	d S_t =  (\alpha + \beta \, \sigma_t^2) S_t \, dt + \sigma_t S_t \, dB_t,  \quad      &  S_t \in \RBB_+, \\
  	d \sigma_t = \sigma_t (\omega - \theta \, \sigma_t) \, dt + \xi \sigma_t \, dW_t,  \quad & \sigma_t \in \RBB_+, \\
  	dB_t \, dW_t = \rho \, dt.          
  \end{array}  \right.  \label{eq:XGBM1}  
  \aeb  
  Here $(B_t,W_t)$ are correlated Brownian motions with correlation $\rho$. Also \newline $(\alpha,\beta,\omega,\theta,\xi,\rho)$ are constant parameters of the model, meant to be estimated from financial time series. We generally assume throughout that ($\omega,\theta,\xi^2)>0$, 
  although we sometimes admit $\omega = \theta = 0$ or even $\theta < 0$.
  
   We also develop
  a risk-neutral (Q-measure) version of the model, which is used for option valuation. 
  In that one, $\alpha$ becomes a cost-of-carry: $\alpha \ra (r-q)$, where $r$ is an interest rate and $q$ is a dividend yield. In addition, 
   $\beta \ra 0$, and
  $\omega \ra \omega_Q$, (a possibly different parameter) while the other parameters
  $(\theta,\xi,\rho)$ remain identical to their P-measure values. 
  
  An attractive feature of the model is that the stochastic volatility $\sigma_t$ is driven by a geometric Brownian motion (GBM)-type noise. GBM volatility seems to be favored by time series analysis over the ``square-root" variance noise of popular affine-type diffusion models, such as 
  the Heston '93 model.\footnote{See, for example \cite{cjm:2010} and \cite{tegpou:2018}.} Extended with the drifts shown, I call it the `Extended GBM' model, or XGBM for short.
  
  \newpage
  
   As we show, the model admits an exact solution for the transition density, vanilla option values,
   and some other quantities of interest. In these solutions, we find two qualitatively
   different regimes:   
   
   \begin{itemize}
   	\item Case 1: \quad $\Smallfrac{1}{2} \xi^2 \le \omega < \infty$,
   	\item Case 2: \quad $0 \le \omega < \Smallfrac{1}{2} \xi^2$.
   \end{itemize}
In Case 1, the driving volatility process is mean-reverting and has a stationary density. 
In Case 2, with probability 1, as $t$ grows large $\sigma_t \ra 0$, similar to the 
SABR model. Indeed, the lognormal SABR model is a special case of (\ref{eq:XGBM1})  
when the drifts are absent. 
   
   \Pbreak
   However, a weakness of the SABR model is its lack of
   a stationary density for the volatility. Thus (under Case 1), we have here what may represent the first example of an SV model combining: GBM-type volatility noise, a stationary volatility density,  and exact solutions. 
   
   \Pbreak
   Numerically, we find option values (and
   thus implied volatilities) continuous vs. $\omega$, including at the
   borderlines case: $\omega = \Smallfrac{1}{2} \xi^2$.
   
   \Pbreak
   In brief, the problem is solvable because it is reducible to the evolution problem for a 1D 
   diffusion operator admitting a spectral expansions in terms of confluent hypergeometric functions $M(a,b,x)$ and $U(a,b,x)$. The other special function that will appear frequently is
   \[ F(a,b,c;z) = \sum_{n=0}^{\infty}\frac{(a)_n (b)_n}{(c)_n} \frac{z^n}{n!},  \] 
   the Gauss hypergeometric function. The $(a)_n$ etc are Pochhammer symbols. All these functions are
   built-in in Mathematica, where everything is implemented.
   
   \Pbreak   
    Structurally, the model is similar to the one-factor short-term
   interest rate model, $d r_t = r_t (\omega - \theta \, r_t) \, dt + \xi r_t \, dW_t$,
   originally due to Merton and which I fully solved in \cite{lewis:1998} (reprinted in \cite{lewis:2016}).
   The solvability of Merton's model suggests, but does not prove, that associated 2D
   SV models  such as (\ref{eq:XGBM1}) are also solvable. Another suggestion that my XGBM SV model
   might be solvable is the discussion found in \cite{hlab:2009} (Sec. 9.5). 
   
   As it turns out, moving from the interest rate model to the associated stochastic volatility model is 
   tricky. One reason is the ``reduced" PDE coefficients become complex-valued,
   which complicates the associated spectral expansion. Another reason is that
   the ``fundamental transform" needed for option valuation (using the language of \cite{lewis:2000a}) requires a regularization for Case 2. All that is explained below; key formulas are boxed.
   
   \newpage

\section{The Joint Transition Probability Density}
 With $(X_t,Y_t) \equiv (\log S_t, \sigma_t)$ the system (\ref{eq:XGBM1}) reads: 

\bea   \left \{ 
\begin{array}{ll}
	d X_t =  [\alpha + (\beta- \smallfrac{1}{2}) \, Y^2_t]  \, dt + Y_t \, dB_t, 
	        \quad    &    X_t \in \RBB, \\
	d Y_t = (\omega Y_t - \theta \, Y^2_t) \, dt + \xi \, Y_t \, dW_t,  \quad & Y_t \in \RBB_+, \\
	dB_t \, dW_t = \rho \, dt.          
\end{array}  \right.  \label{eq:XGBM2}  
\aeb

We seek the bivariate transition probability density $p(t,x',y'|x,y)$ defined by $p(t,x',y'|x,y) \, dx' dy' = \Pr{X_t \in dx', Y_t \in dy' | X_0 = x, Y_0 = y}$. The initial condition is $p(0,x',y'|x,y) = \delta(x' - x) \,\delta(y' - y)$, using the Dirac delta. All parameters are real; assume $(\omega,\theta) \ge 0$.  Using subscripts for derivatives, the corresponding Kolmogorov backward PDE problem is

\bea   \left \{ 
\begin{array}{l}
	p_t = \frac{1}{2} y^2 (p_{xx} + 2 \rho \xi p_{xy} + \xi^2 p_{yy}) + [\alpha + (\beta-\frac{1}{2}) y^2] p_x +  (\omega y - \theta y^2) p_y, \\
	p(0,x',y' | x, y) = \delta(x-x') \, \delta(y-y') .
\end{array}  \right.  \label{eq:KBE} 
\aeb 
The system (\ref{eq:XGBM2}) is a MAP (Markov Additive Process), where $X_t$ is the additive component and $Y_t$ is the Markov component. This is also seen in the $x$-independence in the coefficients of (\ref{eq:KBE}). The MAP property implies \newline $p(t,x',y'|x,y) = p(t,x' - x,y'|0,y)$ which in turn implies the existence of a Fourier representation:

\be  p(t,x',y'|x,y) = \int_{-\infty}^{\infty} \e^{-\i z (x'-x)} \, \Phi(t,z, y, y') \, \frac{dz}{2 \pi},  \label{eq:prep} \eb
for a characteristic function $\Phi$ to be determined below. 

In the Heston '93 model (also a MAP), $\Phi$ is found in terms of Bessel functions. Here the situation is not quite so simple: we'll find that  $\Phi$ itself requires an integration using confluent hypergeometric  functions $M(a,b,x)$ and $U(a,b,x)$. The additional integration means that numerical evaluations of the
exact formulas will take  an order of magnitude more computer time than Heston. Pure numerics, such as a PDE approach, will be similar in the two models. 

\newpage

\subsection{Reduction to an auxiliary problem for $G(\tau,y,y')$} \label{sec:reduction}
Because of the MAP property it suffices to solve (\ref{eq:KBE}) with $x' = 0$. Let $\Phi(t,y,y')= \int \e^{-\i z x} p(t,0,y'|x,y) dx$, suppressing the display of the $z$ dependence. Applying $\int \e^{-\i z x} (\cdots) dx$ to both sides of (\ref{eq:KBE}) and using parts integrations (the boundary terms vanish) yields:

\bea   \left \{ 
\begin{array}{l}
	\Phi_t = \frac{1}{2}  \xi^2 y^2 \Phi_{yy} + b(y;z) \Phi_y - c(y;z) \Phi, \\
		\Phi(0,y,y') = \delta(y-y'),
\end{array}  \right.  \label{eq:KBE2} 
\aeb 
\[ \mbox{where} \quad b(y;z) =  \omega y - (\theta - \i z \rho \xi)  y^2 \,\, \mbox{and} 
 \,\, c(y;z) = -\i z \alpha + [\Smallfrac{1}{2} z^2 - \i z (\beta - \Smallfrac{1}{2})] y^2.\]
 
 \Pbreak
 Letting $\Phi(t,y,y') = \e^{\i z \alpha t} G(\tau,y,y';z)$, where $\tau = \Smallfrac{1}{2} \xi^2 t$ yields
 
 \bea   \left \{ 
 \begin{array}{l}
 	G_{\tau} = y^2 G_{yy} + (\tilde{\omega} y - \theta_z^- y^2) G_y - c^+_z(\beta) y^2 G, \\
 	G(0,y,y') = \delta(y-y'),
 \end{array}  \right.  \label{eq:Greenfunc0} 
 \aeb 
 \[ \mbox{introducing} \quad \tilde{\omega} = \Smallfrac{2 \omega}{\xi^2}, \quad 
  \theta_z^- = \Smallfrac{2}{\xi^2}  (\theta - \i z \rho \xi), \quad \mbox{and}
 \quad c^+_z(\beta) =  \Smallfrac{2}{\xi^2} [\Smallfrac{1}{2} z^2 - \i z (\beta - \Smallfrac{1}{2})].\]
 The $\pm$ superscripts on $\theta_z$ and $c_z$ distinguish them from similar
 expressions introduced later where the sign of $z$ is flipped from the convention here.
 The mnemonic is that $\theta_z^-$ indicates that $z$ occurs as $-\i z$, so the sign is \emph{minus}.
 That is, $\theta_z^{\pm} \equiv \Smallfrac{2}{\xi^2}  (\theta \pm \i z \rho \xi)$. The
 sign on $c^+_z(\beta)$ refers to the risk-neutral case where $\beta = 0$; i.e.,
 $c^{\pm}_z(0) = (z^2 \pm \i z)/\xi^2$.
  
 \Pbreak
 As we see, starting from real $(\omega,\theta) \ge 0$,  and regardless of the $\pm$
 superscripts, the reduced problems have real
 $\tilde{\omega} \ge 0$, while $(\theta_z,c_z) \in \CBB$. If the Fourier inversion is
 performed along the real $z$-axis, and of course $\beta$ is real, we also have
  $\Re \, \theta_z \ge 0$ and $\Re \, c_z \ge 0$.  In terms of $G$, (\ref{eq:prep}) reads:
 \be  p(t,x',y'|x,y) = \int_{-\infty}^{\infty} \e^{-\i z (x'-x - \alpha t)} 
 \, G(\Smallfrac{1}{2} \xi^2 t, y, y';z) \, \frac{dz}{2 \pi},  \label{eq:soln} \eb
 We call $G$ the (auxiliary model) Green function, although we may also refer to its Laplace
 transform as a Green function.
 
\newpage

\newpage

\subsection{Spectral expansion for $G$ (overview)}

Suppressing the $z$-dependence, $G(\tau,y,y')$ has a Laplace transform with respect to $\tau$,
call it $\hat{G}(s,y,y')$, where $s$ is the transform variable. When we
invert the transform via a Bromwich contour in the complex $s$-plane, we'll
discover various singularities which are associated to the spectrum.
See Fig. \ref{fig:XGBMInversionContour} for a typical case.

The simplest situation has real parameters,  $(\omega,\theta,c) \ge 0$.
Then, we'll find that  $\hat{G}(s,y,y')$ has 

\begin{enumerate}
	\item A branch cut singularity at $s = s_c \equiv -\Smallfrac{1}{4}(\omega-1)^2$.
	\item A set of poles (or sometimes an empty set) at $s_n, (n=0,1,\cdots,n_{max})$. These poles are
	are a finite set of non-positive reals,  lying in $s \in (s_c,0)$.
\end{enumerate}
The net result of the Laplace inversion is a spectral expansion of the form:
\be	G(\tau,y,y') =
1_{pc} \sum_{n=0}^{n_{max}} \phi_{n}(y,y') \, \e^{s_n \tau} 
+  \e^{s_c \tau} \,
\int_{0}^{\infty} \phi_{\nu}(y,y') \, \e^{-\nu^2 \tau} \, d \nu,  	\label{eq:Gqualitative} 
\eb
The specifics are given in Sec. \ref{sec:Gresult} and developed in detail in Appendix A..

\Pbreak
The poles generate the discrete spectrum and the discrete sum term in (\ref{eq:Gqualitative}). 
Here $1_{pc}$ denotes a ``pole condition", which is a criterion on the parameters for the
discrete contribution to appear at all. This term may be absent.

\Pbreak
The branch cut singularity marks the right edge of the
$s$-plane interval $(-\infty,s_c)$, which is the continuous spectrum, leading to the integral term in (\ref{eq:Gqualitative}). This term is always present.

\Pbreak
As we saw in Sec. \ref{sec:reduction}, for XGBM we need $\omega \ge 0$,
but $(\theta,c) \ra (\theta_z,c_z)$, complex functions of the Fourier parameter $z$. 
In that case, the general form (\ref{eq:Gqualitative}) still holds, but the structure of the spectrum is altered. 
A continuous spectrum remains along  $(-\infty,s_c)$. However, the poles generally leave the
real $s$-axis, as illustrated in Fig. \ref{fig:XGBMInversionContour}.
The figure illustrates two poles, shown as the `crossed-circles'. 

\Pbreak
Indeed, since (complex) $z$ varies continuously under a Fourier inversion, we have \emph{pole trajectories} $s_n(z)$. Examples are shown in Fig. \ref{fig:SpectrumTrajectories} for a case with (initially) three poles. The illustrated trajectories start in the real interval $(s_c,0)$ as in the real parameter case, then
leave the real $s$-axis, and finally return to touch the branch cut -- at which point they leave the
discrete spectrum. In turn, this leads to other complications which
are seen in Fig. \ref{fig:TransitionPt} and discussed later. Overall, the complex parameter case
is, well, complex!

\newpage

\subsection{Spectral expansion for $G$ (result)} \label{sec:Gresult}

As shown, the transition density we seek is found in terms of a Green function for an auxiliary 1D PDE with (two) complex coefficients. Simplifying the notation, 
the reduced problem (\ref{eq:Greenfunc0}) reads: 
 
 \bea   \left \{ 
 \begin{array}{l}
 	G_{\tau} = y^2 G_{yy} + (\omega y - \theta y^2) G_y - c y^2 G, \\
 	G(0,y,y') = \delta(y-y').
 \end{array}  \right.  \label{eq:Greenfunc} 
 \aeb 
 As before, real $\omega \ge 0$, while $(\theta,c) \in \CBB$ with $\Re \, c \ge 0$.
 Solving (\ref{eq:Greenfunc}) is the key to the XGBM model. From that solution, all the other results in this article follow. I find: 
 
 \begin{theorem}[auxiliary Green function] \label{thm:Green}
 	Let $\lfloor x \rfloor$ denote the greatest integer in $x$, $(a)_n = a (a+1) \cdots (a+n-1)$
 	is the Pochhammer symbol, and the further notations
 	\[ \gamma = -\Smallfrac{1}{2} (\theta - \sqrt{\theta^2 + 4 c}), \quad R = \theta + 2 \gamma, 
 	\quad  \psi = \frac{\omega \gamma}{R}, \quad r_n = -(n+\psi),
 	\]
 	\[ s_n = (n+\psi)^2 + (1-\omega)(n+\psi), \,\, b_n = \omega + 2 r_n,  \,\, a_{\nu} = \Smallfrac{1}{2}(1-\omega) + \psi + \i \nu, \,\, b_{\nu} = 1 + 2 \i \nu. \]
 	\newline       
 	Assume real $\omega \ge 0$, while $(\theta,c) \in \CBB$ with  
 	$\Re \, c \ge 0$. Then, the spectral representation solution to PDE problem $($\ref{eq:Greenfunc}$)$ is given by
 	\begin{empheq}[box=\fbox]{align}
 	&G(\tau,y,y'; \omega,\theta,c) = \nonumber \\
 	&\,\,	R \, (R y')^{\omega-2} \e^{-\gamma y} \e^{-(\theta+\gamma)y'} 
 	\times 1_{\{\frac{\omega-1}{2} > \Re \psi\}}
 	\sum_{n=0}^{\lfloor (\omega-1)/2 - \Re \psi \rfloor} \phi_{n}(y,y') \, \e^{s_n \tau}  \nonumber \\
 	&+ (y')^{\omega-2} (y y')^{(1-\omega)/2} \, \e^{-\gamma y} \e^{-(\theta+\gamma)y'}  
 	\e^{-(1-\omega)^2 \tau/4} \,
 	\int_{0}^{\infty} \phi_{\nu}(y,y') \, \e^{-\nu^2 \tau} \, d \nu,   \label{eq:Gtotal} 
 	\end{empheq}
     where
 	\begin{empheq}{align*}
 	\phi_{n}(y,y') &= \frac{(\omega-1-2n-2\psi) (b_n)_n}{n! \, \Gamma(b_n) } 
 	\,  (R^2 y y')^{r_n} \, M(-n,b_n, R y) \, M(-n,b_n,R y') 
 	\end{empheq}
 	and
 	\be \phi_{\nu}(y,y') = \frac{1}{2 \pi}  \frac{\Gamma(a_{\nu}) \Gamma(a_{-\nu})}{\left| \Gamma(2 i \nu) \right|^2} \,
 	(R^2 y y')^{\i \nu} \, U \left( a_{\nu}, b_{\nu}, R y \right) \, U \left( a_{\nu}, b_{\nu}, R y' \right).  \nonumber \eb
 	\label{prop:ae2}
 \end{theorem}
 Proof: In outline, we construct the solution by Laplace transform  of (\ref{eq:Greenfunc}) with respect to $\tau$. This leads to an ODE reducible to Kummer's differential equation, with Laplace transform parameter
 $s$. The ODE solution 
 $\hat{G}(s,y,y')$ is constructed; we perform the Laplace inversion by a Bromwich contour, which yields a first solution for $G$. Finally the spectral representation (\ref{eq:Gtotal}), which is a second solution
 for $G$, is found  by applying the Residue Theorem to the Bromwich inversion. Details are found in Appendix A.

\newpage

\subsection{Summary for the transition density}

To summarize at this point, the transition density for the P-model XGBM system (\ref{eq:XGBM2}) is given by 

\begin{empheq}[box=\fbox]{align}
	&p_{XGBM}(t,x',y'|x,y) = \int_{-\infty}^{\infty}  \e^{-\i z (x'-x - \alpha t)} \, 
	G \left( \Smallfrac{1}{2} \xi^2 t, y, y'; \tilde{\omega},\theta_z^-,c_z^+ \right)  \, \frac{dz}{2 \pi},  \label{eq:solnfinal}  \\
	\quad &\mbox{where} \,\, \tilde{\omega} = \Smallfrac{2 \omega}{\xi^2}, \quad 
	\theta_z^- = \Smallfrac{2}{\xi^2}  (\theta - \i z \rho \xi), \quad c_z^+ =  \Smallfrac{2}{\xi^2}   
	\left(  \Smallfrac{1}{2} z^2 - \i z (\beta - \Smallfrac{1}{2}) \right), \nonumber \\
	& \mbox{and} \,\, G(t,y,y'; \omega, \theta,c) \,\, \mbox{is given at} \,\, (\ref{eq:Gtotal}). \nonumber 
\end{empheq}

\vspace{-30pt}
\Pbreak
For numerics, it is more efficient to use

\begin{empheq}{align*}
	p_{XGBM}(t,x',y'|x,y) = \int_{0}^{\infty} \Re \left\{ \e^{-\i z (x'-x - \alpha t)} \, 
	G \left( \Smallfrac{1}{2} \xi^2 t, y, y'; \tilde{\omega},\theta_z^-,c_z^+ \right) \right\} \, \frac{dz}{ \pi}.  
\end{empheq}

\subsection{Spectral expansion for $G$ (special cases)}

\subsubsection{The stationary limit.} Suppose $\omega > 1$, $\theta > 0$, and $c=0$ $\Rightarrow \gamma = \psi = 0$. In the large $\tau$ limit, (\ref{eq:Gtotal}) yields 
 \be \Psi(y') \equiv \lim_{\tau \rightarrow \infty} G(\tau,y,y') = 	
       \frac{\theta \, (\theta y')^{\omega-2}}{\Gamma(\omega-1)} \, \e^{-\theta y'}, \quad (\omega > 1). \label{eq:stationayvol} \eb
 This is the stationary density for the (standardized) stand-alone volatility process; i.e., when $\xi^2=2$. When $0 \le \omega \le 1$, then (\ref{eq:Gtotal}) still holds without a stationary limit. In that case, the $Y_t$ particle mass eventually accumulates arbitrarily close to $Y = 0$.   
    For general $\xi^2$, (\ref{eq:stationayvol}) holds with 
 \be (\omega,\theta) \ra (\tilde{\omega},\tilde{\theta}) 
    \equiv \frac{2}{\xi^2}(\omega,\theta) \,\, \mbox{and} \,\,
     \Ex{\sigma_{\infty}} = \int y \, \Psi(y) \, dy = 
     \frac{\tilde{\omega}-1}{\tilde{\theta}}, \,\, (\tilde{\omega}>1).    \label{eq:sigmabar} \eb

 \subsubsection{The norm-preserving case: $\bm{c=0}$.} Suppose $(\omega,\theta) > 0$ in (\ref{eq:Greenfunc}) are real and $c=0$. With no killing, the Green function should be norm-preserving on $\RBB_+$: $\int_0^{\infty} G(t,y,y') \, dy'=1$. 
 Let's check that for the sub-case $\omega > 1$.
 
 \Pbreak
 With $c=0$ and $\theta > 0$, then $\gamma = \psi = 0$. Then, using the stationary
 density from (\ref{eq:stationayvol}), Theorem 1 reads:
  \begin{empheq}{align}
 	&G(\tau,y,y') =  1_{\{ \omega > 1\}} \left[ \Psi(y') 
 	 + \theta \, (\theta y')^{\omega-2} \e^{-\theta y'} 
 	\sum_{n=1}^{ \lfloor (\omega-1)/2 \rfloor} \phi_{n}(y,y') \, \e^{s_n \tau} \right]  \nonumber \\
 	&\quad +  (y)^{(1-\omega)/2} 
 	  (y')^{-3/2 + \omega/2} \, \e^{-\theta y'}  \e^{-(1-\omega)^2 \tau/4} \,
 	\int_{0}^{\infty} \phi_{\nu}(y,y') \, \e^{-\nu^2 \tau} \, d \nu,  	\label{eq:Gtotalspecial} 
 \end{empheq}
 where
 \begin{empheq}{align*}
 	\phi_{n}(y,y') &= \frac{(\omega-1-2n) (b_n)_n}{n! \, \Gamma(b_n) } 
 	\,  (\theta^2 y y')^{-n} \, 	M(-n,b_n, \theta y) \, M(-n,b_n,\theta y') 
 \end{empheq}
 and
 \be \phi_{\nu}(y,y') = \frac{1}{2 \pi}  \frac{\Gamma(a_{\nu}) 
 	\Gamma(a_{-\nu})}{\left| \Gamma(2 i \nu) \right|^2} \,
 (\theta^2 y y')^{\i \nu} \, U \left( a_{\nu}, b_{\nu}, \theta y \right) \, 
 U \left( a_{\nu}, b_{\nu}, \theta y' \right)  \nonumber \eb \label{prop:aespecial}
 now with
 \[ s_n = n (n+1-\omega), \quad b_n = \omega - 2 n, \quad
   a_{\nu} = \Smallfrac{1}{2}(1-\omega) + \i \nu, \quad b_{\nu}=1+ 2 \i \nu \]
   We are also using the convention that $\sum_n^m (\cdots) = 0$ if $m < n$.
   
   \Pbreak
  Recall we suppose $\omega > 1$, so the stationary density exists. Since $\int \Psi(y') \, dy'=1$, the
   remaining terms must have a zero $(y')$-integral. Indeed this is true because, 
   
\[ (i) \quad \int (y')^{-3/2 + \omega/2 + \i \nu} \e^{-y'} U(a_{\nu},b_{\nu}, y') \, dy' = 0, \]
    which follows from taking the parameters $\psi \ra 0$ and $\zeta=1$ in the integral (\ref{eq:SlaterApplied}) below, and
 \[ (ii)  \int (y')^{\omega - 2 - n } \e^{-y'} M(-n,\omega - 2 n, y')\, dy'  = 0, 
    \quad \mbox{for} \quad n = 1,2, \cdots, \lfloor \omega-1 \rfloor. \]
    Here (ii) follows from the relation        
  \be \int (y')^{\omega - 2 - n } \e^{-y'} M(-n,\omega - 2 n, y') \, dy'
    =
   \frac{(\omega - 2 \, n)}{(\omega-1-n) \, \Gamma(1-n)}, \quad (\omega > 1 + n), \eb
   which we found from Mathematica, where $n$ is arbitrary  real, 
   and then specializing to $n=1,2,\cdots, \lfloor \omega-1 \rfloor$. \qed
   
   \pbold{Remarks for the sub-case: $0 \le \omega < 1$.} Norm preservation for this
   sub-case is best shown with the fundamental transform $H$ developed later in Sec. \ref{sec:case2}.
   With that, the property amounts to showing that $H(T,y;z=0)=1$. But, referring to
   (\ref{eq:Hinf}), it is easy to see that $h(y;z=0)=1$ and then $H(T,y;z=0)=1$
   is immediate from (\ref{eq:HCase2})-(\ref{eq:Hreg}).

    \subsubsection{The SABR model limit: no drifts} Drop the drifts
    in (\ref{eq:XGBM1}) and you have the (lognormal) SABR model. Then (\ref{eq:Gtotal}) 
    yields previously
    known SABR results. Details are found in Appendix B.

 \newpage
 
 \section{Risk-neutral evolution}
 Given a P-measure diffusion, to avoid arbitrage opportunities, options must be valued under an equivalent Q-measure diffusion. The usual short-hand trick to finding it are Girsanov substitutions: $dB_t \ra dB^*_t - \lambda^e_t$ and $dW_t \ra dW^*_t - \lambda^v_t$
 in (\ref{eq:XGBM1}). Here $(B^*_t,W^*_t)$ are Q-Brownian motions and $(\lambda^e_t,\lambda^v_t)$ are, respectively,  market prices of equity and volatility risk.
 
  We assume a world with a constant ``cost-of-carry" $b = r - q$, where
 $r$ is a short-term interest rate, and $q$ is a constant dividend yield thrown off by the asset.
 Then, given our
 parametrization in (\ref{eq:XGBM1}), the absence of arbitrage requires 
  \[ \lambda^e_t = \frac{\alpha + \beta \sigma^2_t - (r-q)}{\sigma_t} \]
  In principle, any functional choice $\lambda^v_t = \lambda^v(t,S_t,\sigma_t)$ that preserves the inaccessibility of
  all the spatial boundaries will bring closure to the model and satisfy the  ``no-arbitrage" principle: P-Q equivalence
  as measures. In practice, one wants to preserve `closed formness'. Indeed, a convention in financial modelling (although not required by the general theory) is to
 arrange both the P and Q evolutions to have similar parameterizations. This allows a successful
 solution method for one to work for the other. In this spirit, our
 simple choice here takes $\lambda^v$ a constant.  With that,  
  
   \bea  \mbox{under Q:} \quad \left \{ 
  \begin{array}{l}
  	d S_t = (r-q) S_t \, dt + \sigma_t S_t \, dB^*_t,  \\
  	d \sigma_t = \sigma_t (\omega_Q - \theta \, \sigma_t) \, dt + \xi \sigma_t \, dW^*_t,  \\
  	dB^*_t \, dW^*_t = \rho \, dt,          
  \end{array}  \right.  \label{eq:XGBM3}  
  \aeb
  where $\omega_Q = \omega - \lambda^v \xi$, and now (\ref{eq:XGBM1}) and  (\ref{eq:XGBM3}) together complete the model. 
  
  To simplify the notations, let us agree that all the evolutions and Brownian motions in this  section are `under Q'. Then we have, equivalently, the Q-model:

  \bea    \left \{ 
  \begin{array}{ll}
  	d X_t = (r - q - \frac{1}{2} \, Y^2_t)  \, dt + Y_t \, dB_t,   \\
  	d Y_t = (\omega_Q Y_t - \theta \, Y^2_t) \, dt + \xi \, Y_t \, dW_t,  \\
  	dB_t \, dW_t = \rho \, dt.          
  \end{array}  \right.  \label{eq:XGBM4}  
  \aeb
 
  \newpage
  
  \subsection{The issue of martingality} \label{sec:martingality}
  The XGBM model nests the lognormal SABR model as a special case: see
  (\ref{eq:SABR}) in Appendix B. Recall that, under the lognormal SABR
  model, the stock price process can suffer `loss of martingality' due to the
  correlation parameter $\rho$. Specifically, $S_t$ is: (i) a true
  martingale when $-1 \le \rho \le 0$, and (ii) a strictly local martingale when
  $0 < \rho \le 1$. Since every local martingale that is bounded from below is a 
  supermartingale: $\Esub{S_0,\sigma_0}{S_t} < S_0$ (when $\rho>0)$. 
  
  Indeed, when $\rho > 0$,
  I calculate explicitly in \cite{lewis:2016} the martingale defect. For the full XGBM
  model, correspondingly: when is the discounted stock price process 
  $\tilde{S}_t \equiv \e^{(q-r)t} S_t$ a true martingale? Equivalently, fixing a forward settlement date  $T$, when is the forward price process $F_{t,T} = \e^{(r-q)(T-t)}S_t$ a true martingale?  Suppressing $T$, the evolution for that is:
  
    \bea  \mbox{under Q:} \quad \left \{ 
  \begin{array}{l}
  	d F_t = \sigma_t F_t \, dB^*_t,  \\
  	d \sigma_t = \sigma_t (\omega_Q - \theta \, \sigma_t) \, dt + \xi \sigma_t \, dW^*_t,  \\
  	dB^*_t \, dW^*_t = \rho \, dt,          
  \end{array}  \right.  \label{eq:XGBMF}  
  \aeb  
  A prescription for answering this question is given in \cite{lewis:2000a},\cite{lewis:2016}
  for the general stochastic volatility evolution 
  $d\sigma_t = b(\sigma_t) \, dt + a(\sigma_t) \, dW_t$.
  To determine if this process is martingale-preserving (for the forward),
  introduce the \emph{auxiliary} volatility process $\{\hat{\sigma}_t\}$:
  
  \[ d \hat{\sigma}_t = [b(\hat{\sigma}_t)+ \rho \, \hat{\sigma}_t a(\hat{\sigma}_t)] dt + a(\hat{\sigma}_t) dW_t. \]
  Then, the forward $F_t$ suffers a loss of martingality if and only if the auxiliary process can
  `explode'; i.e., reach $\hat{\sigma} = +\infty$ in finite time with strictly positive probability. From
  (\ref{eq:XGBMF}), for the case under consideration,
    \[ d \hat{\sigma}_t = (\omega_Q \, \hat{\sigma}_t - \delta \, \hat{\sigma}^2_t) dt + 
    \xi \hat{\sigma}_t dW_t, \quad  
    \mbox{where} \quad  \delta =  \theta - \rho \, \xi. \]
  Then, by applying the \emph{Feller explosion criteria} (details in \cite{lewis:2000a}), one
  	can establish that $\hat{\sigma}_t$ can explode if and only if $\delta < 0$; i.e.,
  	$\rho > \theta/\xi$. 
  	
  	Even better, when	$\rho > \theta/\xi$, one can find the \emph{martingale defect} by,
  	equivalently, finding the \emph{absorption-at-zero probability} $A(x,t)$ for the inverse process
  	$X_t = 1/\hat{\sigma}_t$. From \Ito,
  	\be dX_t = (\alpha X_t + \delta) dt + \xi X_t \, dW_t, \quad
  	       \mbox{where} \quad \alpha = \xi^2 - \omega_Q. \label{eq:auxX} \eb
  	 With $\tau_0$ the first time the $X$-process hits zero, $A(x,t)$ is defined by
  	 \[  A(x,t) = \Pr{X(\tau_0) \le t| X_0 = x }.   \]
  	 Then, the martingale defect is given by 
  	 $\Esub{S_0,\sigma_0}{S_t/S_0} = 1 - A(1/\sigma_0,t)$, with an explicit formula found 
  	 in Appendix C.
  	
  	In summary, the XGBM model, parametrized at (\ref{eq:XGBM3})
  	suffers a loss of martingality for the forward only when $\rho$ is found in the interval
  	$\theta/\xi < \rho \le 1$.
  	For (broad-based) equity applications, the condition is typically harmless:  
    one will usually estimate both $\rho < 0$ (negative option `skews') and $\theta > 0$ (volatility
  	has a stationary density). But for other applications (possibly with 
  	positive option skews or negative $\theta$'s), it is an issue to keep in mind.

  \newpage

  \section{Option valuation -- Case 1: $\omega_Q \ge \xi^2/2$}
  Let $q(t,x',y'|x,y)$ denote the XGBM transition density under the Q-evolution. From (\ref{eq:solnfinal}),
  with $\alpha \ra (r-q)$, $\beta \ra 0$, and taking dummy $z \ra -z$, we have  
  
  \begin{empheq}{align}
  	&q(t,x',y'|x,y) = \int_{-\infty}^{\infty}  \e^{\i z (x'-x - (r-q) t)} \, 
  	G \left( \Smallfrac{1}{2} \xi^2 t, y, y'; \tilde{\omega},\theta_z^+,c_z^- \right)  \, \frac{dz}{2 \pi},  \label{eq:solnfinalq}  \\
  	\quad &\mbox{where} \,\, \tilde{\omega} = \Smallfrac{2 \omega_Q}{\xi^2}, \quad 
  	\theta_z^+ = \Smallfrac{2}{\xi^2}  (\theta + \i z \rho \xi), \quad c_z^- =  \Smallfrac{1}{\xi^2}   
  	\left( z^2 - \i z \right). \label{eq:qparms}
    \end{empheq}
Here $G(t,y,y'; \omega, \theta,c)$ is again given at (\ref{eq:Gtotal}). Let $V(x,y,T)$ denote
the time-0 value of a Euro-style option with time-$T$ {\underline{bounded}} payoff function $w(x)$. 
Then, since $V$ is the time-0 discounted expected value of the payoff under the
$Q$-evolution, we have generally

\begin{empheq}{align}
 & V(x,y,T) = \frac{\e^{-r T}}{2 \pi} \nonumber \\
  &\times \smallint_{-\infty}^{\infty} \smallint_{0}^{\infty} \smallint_{-\infty}^{\infty} 
 \e^{\i z (x'-x - (r-q) T)} w(x') \, 
G \left( \Smallfrac{1}{2} \xi^2 T, y, y'; \tilde{\omega},\theta_z^+,c_z^- \right)  \, dz \,dy' \, dx'. \label{eq:optionval}
\end{empheq}
 While (\ref{eq:optionval}) offers a quasi-analytic solution, it has an embarrassing number of
  integrations: 4(!), counting the integration for  $G$ itself. We'll eliminate two.
 
  \Pbreak
Proceeding formally, suppose it's legitimate to do the
$y'$ integral in (\ref{eq:optionval}) \emph{first}. Then, \emph{if the following integral
exists}, define a \emph{fundamental transform} 

\be \mbox{?} \quad H(T,y;z) \equiv \int_{0}^{\infty} 
  G \left( \Smallfrac{1}{2} \xi^2 T, y, y'; \tilde{\omega},\theta_z,c_z \right) \, dy',
    \quad z \in \mathcal{S}_1, \label{eq:H} \eb
 where $\mathcal{S}_1$ is an  analyticity strip for $H$ in the complex $z$-plane. 
 Note that we are now generalizing (\ref{eq:optionval}) by moving the $z$-integration off
 the real $z$-axis.
 
 For options, our preferred strip for Fourier inversions is: $0 < \Im z < 1$. It can be shown that
 our preferred strip is contained within $\mathcal{S}_1$ for XGBM because the model is
 (i)norm-preserving and (ii) martingale-preserving 
 (under the mild restriction $-1 \le \rho \le \min(1,\theta/\xi))$.\footnote{To
 	show that (i) and (ii) lead to analyticity in the preferred strip, see Th. 4.7 
 	in \cite{lewis:2016}.} 
 
 We will show below that the question mark in front of (\ref{eq:H}) can
 be removed for Case 1.  
  Next, in a familiar argument, define the (generalized) \emph{payoff transform} 
 \be \hat{w}(z) = \int_{-\infty}^{\infty} \e^{i z x} w(x) \, dx, \quad z \in \mathcal{S}_w, \label{eq:wtransform} \eb
 with $z$ located in some other analyticity strip $\mathcal{S}_w$. We can always
 find a nice payoff such that these two horizontal strips intersect: $\mathcal{S}_V = \mathcal{S}_1 \cap \mathcal{S}_w \not= \emptyset$. 
 Finally, by the arguments in my article ``A Simple Option Formula for General Jump-diffusion and other Exponential \Levy Processes" (reprinted in \cite{lewis:2016}), (\ref{eq:optionval}) becomes

    \be  V(x,y,T) = \frac{\e^{-r T}}{2 \pi} \int_{\i c -\infty}^{\i c + \infty} 
     \e^{-\i z (x + (r-q) T)} 
   H(T,y;z) \, \hat{w}(z) \ dz, \quad z \in \mathcal{S}_V. \label{eq:optionval3} \eb
In other words, $\i c$ is a point in the $z$-plane marking the intersection of a horizontal integration contour in $\mathcal{S}_V$ with the imaginary  $z$-axis. Again, we can arrange things
so that we are working in the preferred strip: $0 < c < 1$. Reported numerics are found from versions of (\ref{eq:optionval3}). 
For example, with $x = \log S_0$, $X = (r-q)T + \log S_0/K$ with strike price $K$, and
 $c = \Smallfrac{1}{2}$, all reported put option values $P(x,y,T)$ are computed from:

\be  P(x,y,T) = K \e^{-r T} \left( 1 -  \int_{\i c}^{\i c + \infty} 
  \Re \left\{ \e^{-\i z X} 
\frac{H(T,y;z)}{z^2 - \i z} \right\} \, \frac{dz}{\pi} \right). \label{eq:putval} \eb

\pbold{Removing the question mark in (\ref{eq:H}).}  Recall that the SABR model is a special
case of the XGBM model, with the transition density given at (\ref{eq:pSABRalt}) in Appendix B.
It turn out that, for the SABR model, the integral in (\ref{eq:H}) does \emph{not} exist. 

This problem was discussed in \cite{lewis:2016} (Sec. 8.13, pg. 407), but we will recap briefly
here. For SABR, the existence of (\ref{eq:H}) amounts to integrating (\ref{eq:pSABRalt}) from 
Appendix B with respect to
$y'$, at fixed $(\nu,z)$. Extracting the few terms in  (\ref{eq:pSABRalt})
with $y'$ dependence yields the integral
\[        \int_0^{\infty} (y')^{-3/2} e^{-\i z \rho y'/\xi}  K_{\i \nu}(\chi_z y') \, dy',                   \]
which does not exist because the integrand is not integrable 
as $y' \downarrow 0$.\footnote{\label{ft:bes}$K_{\i \nu}(y)
 \sim c_1 y^{-\i \nu} + c_2  y^{\i \nu}$, as $y \ra 0$, where $c_{1,2}$ depend upon $\nu$ but not $z$.} 

Now, for the SABR limit, there is no discrete spectral contribution to $G$
in (\ref{eq:Gtotal}). The lesson from SABR is that, in the general XGBM case, problems may lie with
integrating the continuous spectrum term.  Indeed, extracting all the $y'$ dependencies
from the continuous spectrum term in (\ref{eq:Gtotal}), the corresponding XGBM integral is
 \[   (*) \quad  \int_0^{\infty} (y')^{\tilde{\omega}/2-3/2} 
 e^{-(\theta_z^+ + \gamma_z) y'} (y')^{\i \nu} U(a_{\nu},b_{\nu}, R_z y' ) \, dy'.  \]
 \[ \mbox{Recall} \quad \tilde{\omega} = \Smallfrac{2 \omega_Q}{\xi^2}, \quad 
 \theta_z^+ = \Smallfrac{2}{\xi^2}  (\theta + \i z \rho \xi), \quad 
      c_z^- =  \Smallfrac{1}{\xi^2}   
 \left( z^2 - \i z \right), \]
 \[ \mbox{and so} \quad \gamma_z = -\Smallfrac{1}{2} \left( \theta_z^+ -
   \sqrt{(\theta_z^+)^2 + 4 c_z^-} \right), \quad R_z = \theta_z^+ + 2 \gamma_z = \sqrt{(\theta_z^+)^2 + 4 c_z^-}. 
        \]   
 Now, first consider the integrand as $y' \downarrow 0$. As $x \ra 0$,
 $U(a,b,x) \sim c_1 + c_2 x^{1-b}$, where $c_{1,2}$ are bounded constants independent of $x$.
 Since $b_{\nu} = 1 + 2 \i \nu$,   
 \[  (y')^{\i \nu} U(a_{\nu},b_{\nu}, R_z y' ) \sim c_1 (y')^{\i \nu} +
           c_2   (y')^{-\i \nu}, \]
           much like the SABR case with Bessel functions discussed in footnote \ref{ft:bes}. The
           exponential in harmless at small $y'$, so we see the crux of the
           matter is the term $(y')^{\tilde{\omega}/2-3/2}$, which is integrable
            iff\footnote{Integrability is clear when $\tilde{\omega}>1$. The
            	borderline case $\tilde{\omega}=1$ is not as immediate, but (\ref{eq:SlaterApplied}) below
            shows that (*) also exists when  $\tilde{\omega}=1$ as long as $\nu>0$.}
     
     \begin{empheq}[box=\fbox]{align}      
         \tilde{\omega} \ge 1, \quad (\mbox{i.e.,} \,\,
                 \omega_Q \ge \Smallfrac{1}{2} \xi^2), \quad   \mbox{(Case 1: integrability condition)}
                  \label{eq:integrability} 
     \end{empheq}  
  When the integrability condition holds, so does (\ref{eq:H}). 
   Recall we have seen this condition before at (\ref{eq:stationayvol}) in the P-model: there 
   needed for the existence of a stationary limit of the stand-alone volatility process. In
   most financial applications (with calibrated parameters), the integrability condition will be satisfied.
   Thus, to borrow a characterization from physics, Case 1 is the ``physical case". Case 2 is the
   ``unphysical case" (including SABR). Case 2 is, of course, still mathematically well-defined.
   
   \Pbreak   
        What about $y' \ra \infty$? Since 
       \[ \theta_z^+ + \gamma_z = \Smallfrac{1}{2} \left( \theta_z^+ +
        \sqrt{(\theta_z^+)^2 + 4 \, c_z^-} \right),  \]
        this term will have positive real part in the preferred $z$-plane strip. As the
        remaining integrand terms in (*) have power law behavior as $y' \ra \infty$, the
        net effect is exponential decay in $y'$ at large $y'$, so all is well in that limit.
 
 \Pbreak       
         To summarize at this point, we have the following situation.
         In the risk-neutral model, when the integrability condition (\ref{eq:integrability}) holds, 
        option values are given by (\ref{eq:optionval3}), with $H$ given by (\ref{eq:H}). If the integrability condition does \emph{not} hold
           (Case 2), option values are \emph{still} given by (\ref{eq:optionval3}), 
           but $H(\cdot)$ is \emph{not}
           given by (\ref{eq:H}). Instead, that integral needs to be regularized:
           see Sec. \ref{sec:case2} for details. In this section, we now continue, assuming
           (\ref{eq:integrability}) holds.

  \pbold{Option valuation -- continued}.
   To perform the integration in (*), use a formula from
   Slater (\cite{slater:1960}, (3.2.51)-(3.2.52)):
   
 \begin{empheq}{align*}
  & \int_0^{\infty} \e^{-s t} t^{c-1} U(a,b,t) \, dt  \nonumber  \\ 
 &= \frac{\Gamma(c) \Gamma(1+c-b)}{\Gamma(1+a+c-b)} \times \left \{ 
 \begin{array}{ll}
 	     F(c,1+c-b,1+a+c-b;1-s), \quad &|1-s| < 1, \\
         s^{-c} F \left(a,c,1+a+c-b;1-\frac{1}{s} \right), \quad  &\Re \, s > \Smallfrac{1}{2},                    
 \end{array}  \right. 
\end{empheq}
\be \mbox{for} \,\, \Re c>0, \quad \Re b < \Re c + 1.   \label{eq:SlaterUintegral} \eb  
Here $F(a,b,c;z) = \sum_{n=0}^{\infty}\frac{(a)_n (b)_n}{(c)_n} \frac{z^n}{n!}$ is
the Gauss hypergeometric function. Note that the two conditions on $s$ above
can overlap, allowing both integral relations to
be true at the same time.  
Applying (\ref{eq:SlaterUintegral}) to our problem (*) above yields:
\begin{empheq}{align*}
  \int_0^{\infty} (y')^{\tilde{\omega}/2-3/2+\i \nu} 
	& e^{-(\theta_z^+ +\gamma_z) y'} U(a_{\nu},b_{\nu}, R_z y' ) \, dy' \nonumber  \\ 
	&=\frac{|\Gamma(c_{\nu})|^2}{R_z^{c_{\nu}}\Gamma(\psi)} \times \left \{ 
	\begin{array}{ll}
		F(c_{\nu},c^*_{\nu},\psi_z;1- \zeta_z), \quad &|1 - \zeta_z| < 1, \\
		s^{-c_{\nu}} F \left(a_{\nu},c_{\nu},\psi_z;1-\frac{1}{\zeta_z} \right), \quad  &\Re \, \zeta_z > \Smallfrac{1}{2},                    
	\end{array}  \right. 
\end{empheq}
\be \mbox{where} \quad 
  a_{\nu} = \Smallfrac{1}{2}(1-\tilde{\omega}) + \psi_z + \i \nu, \quad 
  c_{\nu} =- \Smallfrac{1}{2}+\Smallfrac{1}{2} \tilde{\omega}+\i \nu, \quad
 \psi_z = \frac{\tilde{\omega} \gamma_z}{R_z}, \label{eq:SlaterApplied} \eb
 \be \mbox{and} \quad \zeta_z = \frac{1}{2} 
 \left( 1 + \frac{\theta_z^+}{\sqrt{(\theta_z^+)^2 + 4 \, c_z^-}} \right). \label{eq:sdef} \eb
 Note $c^*_{\nu} = - \Smallfrac{1}{2}+ \Smallfrac{1}{2} \tilde{\omega}-\i \nu$ denotes the 
 complex conjugate of $c_{\nu}$ (since $\tilde{\omega}$ is real). The integration
 conditions $\Re c>0$ and $\Re b < \Re c + 1$ from Slater reduce to: $\tilde{\omega}>1$. 
 (Recall in fact $\tilde{\omega}=1$ is OK for Case 1 as long as $\nu \not= 0$).
 The $\zeta_z$-value restrictions can be checked along any putative
 $z$-plane integration contour; I have found the $|1 - \zeta_z|<1$ formulas suffice for all
 the test examples of Sec. \ref{sec:numerics}.
 
 \pbold{The discrete term integration.}
   From (\ref{eq:Gtotal}) and (\ref{eq:H}), the discrete term in the spectral expansion requires  
 \[   (**) \quad  \int_0^{\infty} (y')^{\tilde{\omega}-2+r_n} 
 e^{-(\theta_z^+ +\gamma_z) y'}  M(-n,b_n, R_z y' ) \, dy',  \]
    where $r_n = -(n+\psi)$ and $\psi$ is given at (\ref{eq:SlaterApplied}).
      The integral (**) can be performed using another formula from Slater, (3.2.16):
    \be
     \int_0^{\infty} \e^{-s t} t^{c-1} M(-n,b,kt) \, dt = 
     \Gamma(c) \, s^{-c} F \left(-n,c,b;\frac{k}{s} \right), 
     \label{eq:SlaterMintegral} \eb      
   \[ \mbox{for} \,\, \Re c>0, \,\, \Re s > 0, \,\, \mbox{and} \,\, n=0,1,2,\cdots \]  
   Note that, since $n$ is an integer for our application, $F(-n,c,b,z)$ is a terminating polynomial of $O(z^n)$. Thus,
   \begin{empheq}{align*}
     \int_0^{\infty}  (y')^{\tilde{\omega}-2+r_n} 
   e^{-(\theta_z^+ +\gamma_z) y'} & M(-n,b_n, R_z y' ) \, dy' \\
      &= \Gamma(c_n) \, (R_z \zeta_z)^{-c_n}  
 F \left(-n,c_n,b_n;\frac{1}{\zeta_z} \right),
 \end{empheq}
 \[ \mbox{where} \quad c_n = \tilde{\omega} -1 + r_n, \quad b_n= \tilde{\omega} + 2 \, r_n,
 \,\, \mbox{and $\zeta_z$ is defined at (\ref{eq:sdef})}. \] 
 
 \pbold{Remarks.} Note that the last Slater integral (\ref{eq:SlaterMintegral}) requires
 $\Re \, c > 0$, which translates to $\Re \, c_n > 0$. It can be seen that this condition is
 automatically satisfied here as follows.  First, $\Re \, c_n > 0$ is equivalent to  
 $\tilde{\omega} - 1 > (n + \Re \, \psi)$. But, if $s_n$ contributes to the discrete spectrum, then
 $\tilde{\omega} - 1 > 2(n + \Re \, \psi)$ (see Appendix A or the sum cutoff
in (\ref{eq:Htotal}). When this last inequality is
 satisfied, the l.h.s. will be positive. And certainly if $A > B$ with $A$ 
 positive but otherwise arbitrary, then $A > B/2$.

\newpage

\subsection{Frequently used notations under risk-neutrality}  \label{sec:notations}
Here we collect in one place notation we have introduced and will
frequently refer to subsequently.
\begin{empheq}[box=\widefbox]{align*}	
 &\tilde{\omega} = \Smallfrac{2 \omega}{\xi^2}, \quad 
\theta_z^+ = \Smallfrac{2}{\xi^2}  (\theta + \i z \rho \xi), \quad 
c_z^- =  \Smallfrac{1}{\xi^2}   
\left( z^2 - \i z \right),    \\
&R_z = \sqrt{(\theta_z^+)^2 + 4 \, c_z^-},   \quad
\gamma_z = -\Smallfrac{1}{2} \theta_z^+ + 
\Smallfrac{1}{2} R_z,
\quad 
\zeta_z = \Smallfrac{1}{2} 
\left( 1 + \frac{\theta_z^+}{R_z} \right), \quad
\psi_z = \frac{\tilde{\omega} \, \tilde{\gamma_z}}{R_z},   \\
&a_{\nu} = \Smallfrac{1}{2}(1-\tilde{\omega}) + \psi_z + \i \nu, \quad
b_{\nu} = 1 + 2 \i \nu, \quad
c_{\nu} =- \Smallfrac{1}{2}+\Smallfrac{1}{2} \tilde{\omega}+\i \nu, \\  
&r_n = -(n + \psi_z), \quad s_n(z) = (n+\psi_z)^2 + (1-\tilde{\omega})(n+\psi_z),\\
&b_n= \tilde{\omega} + 2 \, r_n, \quad 
c_n = \tilde{\omega} -1 + r_n.   
\end{empheq}
Note that here and throughout, we frequently suppress $z$-dependencies to ease notations, 
as we have done with $(a_{\nu},r_n,b_n,c_n)$.

\subsection{Summary: fundamental transform for Case 1}

Putting it all together, for $\omega \ge \xi^2/2$, we have the spectral representation:
 \begin{empheq}[box=\widefbox]{align}
 	&\,\, \, H(T,y;z) = \int_0^{\infty} G(\tau,y,y'; \tilde{\omega},\theta_z^+,c_z^-) \, dy' 	\label{eq:Htotal} \\
 	&\,\,= \e^{-\gamma_z y}  \left\{
 	1_{\{\frac{\tilde{\omega}-1}{2} > \Re \psi_z \}} \hspace{-18pt}
 	\sum_{n=0}^{ \lfloor (\tilde{\omega}-1)/2 - \Re \psi_z \rfloor} \phi_{n}( R_z y) \, \e^{s_n(z) \tau} 
 	     	 +   
 	\int_{0}^{\infty} \phi_{\nu}( R_z y) \, 
 	\e^{-(\nu^2 + \frac{1}{4}(1-\tilde{\omega})^2) \tau} \, d \nu \right\}  	\nonumber 
 \end{empheq}

\vspace{-40pt}
 \begin{empheq}{align*}
 	& \mbox{where} \,\, \tau = \Smallfrac{1}{2} \xi^2 T, \\
 &	\phi_{n}(y) = \frac{(\tilde{\omega}-1-2n-2\psi) (b_n)_n}{n!} 
 	  \frac{\Gamma(c_n)}{\Gamma(b_n)}
 	\,\,   \zeta_z^{-c_n}	F \left(-n,c_n,b_n;\frac{1}{\zeta_z} \right) 	
          y^{r_n}	M(-n,b_n, y),  \\ 
 & \phi_{\nu}(y) = \frac{1}{2 \pi} 
  \frac{\Gamma(a_{\nu}) \Gamma(a_{-\nu})|\Gamma(c_{\nu})|^2}{\left| \Gamma(2 i \nu) \right|^2 \Gamma(\psi_z)} \, y^{(1-\tilde{\omega})/2 + \i \nu}
     \, U \left( a_{\nu}, b_{\nu}, y \right)  \\
 	&\quad\quad\quad\quad \times \left \{ 
 \begin{array}{ll}
 	F(c_{\nu},c^*_{\nu},\psi_z;1-\zeta_z), \quad &|1-\zeta_z| < 1, \\
 	\zeta^{-c_{\nu}} F \left(a_{\nu},c_{\nu},\psi_z;1-\frac{1}{\zeta_z} \right), \quad  &\Re \, \zeta_z > \Smallfrac{1}{2},                    
 \end{array}  \right. \\ 
  \end{empheq}
with other notations in  Sec. \ref{sec:notations}. Again, some $z$-dependencies are suppressed.

\newpage

  \section{Option valuation -- Case 2: $0 \le \omega_Q < \xi^2/2$} \label{sec:case2}
  
  To handle this case, we'll adapt the method I used for the SABR model in \cite{lewis:2016}.
  Consider $h(y)$, a putative time-independent solution to the PDE 
  in the top line of (\ref{eq:Greenfunc}). Suppose $h(y) \sim 1 + O(y^{1- \tilde{\omega}})$
  as $y \ra 0$ and $h(y)$ decays at large $y$. Note an initial condition plays no role in this time-independent
  problem. (With $\theta \ra \theta_z$ and $c \ra c_z$,
  we write this solution as $h(y;z)$). With that putative solution, the
  fundamental transform $H(T,y;z)$ can be constructed as
  
  \be H(T,y;z) = h(y;z) +  H_{reg}(\tau,y;z), \quad (\tau = \Smallfrac{1}{2} \xi^2 T).
   \label{eq:HCase2} \eb  	
 Here a \emph{regularized} fundamental transform $H_{reg}(\tau,y;z)$ is defined by
  
  \be  H_{reg}(\tau,y;z) \equiv \int_{0}^{\infty} (1 - h(y';z)) \,
  G \left( \tau, y, y'; \tilde{\omega},\theta_z^+,c_z^- \right) \, dy',
  \quad z \in \mathcal{S}_1. \label{eq:Hreg} \eb
  Let's review why this works. First, both terms on the r.h.s. of (\ref{eq:HCase2})
  manifestly satisfy the PDE associated to the fundamental transform -- the top line of (\ref{eq:Greenfunc}) again. Second, as $T \ra 0$, because $G(\cdots)$ satisfies a Dirac mass initial
  condition, we have $H_{reg}(0,y;z) = 1 - h(y;z)$. Thus, $H(0,y;z) = 1$, which is the correct initial condition for the fundamental transform. Third, the integral in (\ref{eq:Hreg})
  exists because $1 - h(y;z) \sim  O(y^{1- \tilde{\omega}})$ as $y \ra 0$, so the problematic integrand in (*)
  is tamed by this new behavior. Finally, $H(T,y;z)$ correctly decays at large $y$ because both terms on the r.h.s. of (\ref{eq:HCase2}) do.
  
  \Pbreak
  Using the notations in Sec. \ref{sec:notations} and the confluent hypergeometric $U$, I find
  \begin{empheq}[box=\fbox]{align}
  h(y;z) = \e^{-\gamma_z y}  
          \frac{U(\psi_z,\tilde{\omega}, R_z y)}{U(\psi_z,\tilde{\omega}, 0)}, 
                   \label{eq:Hinf} 
   \end{empheq}     
which has the advertised $y$-behaviors.           
  Now  recall 
  \be U(a,b,z) = \frac{\Gamma(1-b)}{\Gamma(a-b+1)} M(a,b,z) 
       + z^{1-b}  \frac{\Gamma(b-1)}{\Gamma(a)} M(a-b+1,2-b,z). \label{eq:Udef} \eb     
 Since $M$ has a Taylor expansion about $z=0$, this implicitly defines coefficients $(A_n,B_n)$ such that
 \[ h(y;z) = \sum_{n=0}^{\infty} A_n(z) \, (R_z y)^n + (R_z y)^{1-\tilde{\omega}}
          \sum_{n=0}^{\infty} B_n(z) \, (R_z y)^n. \]
  I find (again using the Gauss hypergeometric function $F(a,b,c;z)$) that
  \begin{empheq}{align*}
   A_n(z) &= f_n(\psi_z, \tilde{\omega}, \zeta_z -1), \\
   B_n(z) &= \frac{\Gamma(\tilde{\omega}-1) \Gamma(\psi_z - \tilde{\omega} +1)}
              {\Gamma(1-\tilde{\omega}) \Gamma(\psi_z)}  f_n(\psi_z-\tilde{\omega}+1, 2-\tilde{\omega}, \zeta_z -1),  \quad \mbox{where} \\
 f_n(a,b,c) &= c^n \, \frac{F(a,-n,b;-\frac{1}{c})}{\Gamma(1+n)}.                
              \end{empheq}

          Note that, since $f_0=1$, the first sum of $h$ is $1 + O(y)$ and the second sum is
          $O(y^{1-\tilde{\omega}})$; since $\tilde{\omega}<1$, the net effect is  $h(y) \sim 1 + O(y^{1- \tilde{\omega}})$, as promised.
          
          The term-by-term integrals that are now needed in (\ref{eq:Hreg}) can be done
          by the previously given formulas from Slater. The net result is that the previously introduced $\phi_{\nu}(y)$ (for Case 1) generalizes to  $\phi_{\nu,n}(y)$ where
          
          \begin{empheq}{align*}
               	& \phi_{\nu,n}(y) = \frac{1}{2 \pi} 
          	\frac{\Gamma(a_{\nu}) \Gamma(a_{-\nu})|\Gamma(c_{\nu,n})|^2}{\left| \Gamma(2 i \nu) \right|^2 \Gamma(\psi_z)} \, y^{(1-\tilde{\omega})/2 + \i \nu}
          	\, U \left( a_{\nu}, b_{\nu}, y \right)  \\
          	&\quad\quad\quad\quad \times \left \{ 
          	\begin{array}{ll}
          		F(c_{\nu,n},c^*_{\nu,n},\psi_z+n;1-\zeta_z), \quad &|1-\zeta_z| < 1, \\
          		(\zeta_z)^{-c_{\nu,n}} F \left(a_{\nu},c_{\nu,n},\psi_z+n;1-\frac{1}{\zeta_z} \right), \quad  &\Re \, \zeta_z > \Smallfrac{1}{2},                    
          	\end{array}  \right. \\ 
          \end{empheq}       
      In addition, we need to introduce the new function $\psi_{\nu,n}(y)$ where
      
      \begin{empheq}{align*}
      	& \psi_{\nu,n}(y) = \frac{1}{2 \pi} 
      	\frac{\Gamma(a_{\nu}) \Gamma(a_{-\nu})|\Gamma(d_{\nu,n})|^2}{\left| \Gamma(2 i \nu) \right|^2 \Gamma(1-\tilde{\omega}+n+\psi_z)} \, y^{(1-\tilde{\omega})/2 + \i \nu}
      	\, U \left( a_{\nu}, b_{\nu}, y \right)  \\
      	&\quad\quad\quad\quad \times \left \{ 
      	\begin{array}{ll}
      		F(d_{\nu,n},d^*_{\nu,n},1-\tilde{\omega}+\psi_z+n;1-\zeta_z), \quad &|1-\zeta_z| < 1, \\
      		(\zeta_z)^{-d_{\nu,n}} F \left(a_{\nu},d_{\nu,n},1-\tilde{\omega}+\psi_z+n;1-\frac{1}{\zeta_z} \right), \quad  &\Re \, \zeta_z > \Smallfrac{1}{2},                    
      	\end{array}  \right. \\ 
      \end{empheq} 
  These functions use the new notations:
  \[ c_{\nu,n} = n - \Smallfrac{1}{2}+\Smallfrac{1}{2} \tilde{\omega}+\i \nu, \quad (n \ge 1) \] 
  \[ d_{\nu,n} = n + \Smallfrac{1}{2} - \Smallfrac{1}{2} \tilde{\omega}+\i \nu, \quad (n \ge 0). \]
  From these, form
  
  \be \kappa(y;z) \equiv \sum_{n=1}^{\infty} \, A_n(z) \, \phi_{\nu,n}(y) +  \sum_{n=0}^{\infty} 
            \, B_n(z) \, \psi_{\nu,n}(y) \label{eq:kappadef} \eb 
   and obtain finally, for Case 2,
   \begin{empheq}[box=\fbox]{align}
   H(T,y;z) = h(y;z) - \e^{-\gamma_z y} \int_0^{\infty} 
       \kappa_{\nu}(R_z y; z) \, \e^{-(\nu^2 + \frac{1}{4}(1 - \tilde{\omega}^2)) \tau} \, d \nu.  
       \label{eq:HCase2final}
       \end{empheq}
   Don't forget that $\tau = \Smallfrac{1}{2} \xi^2 T$ --  both here and in (\ref{eq:Htotal}).
   Equation (\ref{eq:HCase2final}) is the Case 2 replacement for (\ref{eq:Htotal}); option values
   are computed from this new $H$ using the previous  (\ref{eq:optionval3}) and (\ref{eq:putval}).

  \newpage

   \section{Small and large time asymptotics}
   
   By `time', we refer to the time-to-option-expiration $T$.
   
   \subsection{Small $T$}
   For small $T$, option prices at strike $K$ tend to their parity
   values, and the implied volatility smile, $\sigma_{imp}(T,K,S_0,\sigma_0)$, tends
   to a non-trivial limit. 
   Indeed, it is well-known that the limiting asymptotic smile in general SV models
   does not depend upon the diffusion drifts; it is solely determined
   by the variance-covariance system. In our case, that system is (\ref{eq:SABR}), which is the (lognormal) SABR model. 
   The asymptotic smile formula for that one is well-known. The version I like is\footnote{I am copying in (\ref{eq:smallTIV}) from eq. (12.33) in \cite{lewis:2016} (Ch. 12). It is easily shown to be equivalent
  	to other SABR model sources. For example, to show it is equivalent to (8.32) in \cite{paulot:2010}, identify my $z$ with $-(2 \nu q/\alpha)$ in Paulot and apply some
     routine algebra.} 
   
   \be   \sigma^{imp}_0 \equiv \lim_{T \ra 0} \sigma_{imp}(T,K,S_0,\sigma_0) = 
   \frac{ z \, \sigma_0}{F(z)},
          \quad \mbox{where} \quad z = \left(\frac{2 \xi}{\sigma_0} \right) \log \frac{S_0}{K}, \label{eq:smallTIV} \eb
   \[    \mbox{and} \quad   
   F(z) = 2 \log \left\{ \frac{z/2-\rho + \sqrt{1 - \rho z + z^2/4}}{1- \rho} \right\}. \]

   \subsection{Large $T$}
   \label{sec:largeT}
   
   The large $T$ option behavior is determined from the large $T$ behavior
   of the fundamental transform $H(T,\sigma;z)$.
   
   \pbold{Case 1: $\tilde{\omega} \ge 1$}. In this case, following the method in \cite{lewis:2000a} (Ch. 6),
   one looks for the behavior (as $T$ grows large)  $H(T,\sigma;z) \sim \e^{-\lambda_0(z) T} u_0(\sigma;z)$ in the spectral expansion (\ref{eq:Htotal}). When there is a discrete
   component in that expansion, then $\lambda_0$ is the principal eigenvalue.
   When there is only a continuous spectrum, but still real $\lambda_0 > 0$, a better nomenclature
   for $\lambda_0$  (borrowing from physics)  might be be `mass gap'. In any event, we define $\lambda_0$ by
   \[ \lambda_0(z) \equiv \lim_{T \ra \infty} -\frac{1}{T} \log H(T,\sigma;z). \]

   Tentatively, assume the parameters are such that there is indeed a discrete spectrum,
   and identify $\lambda_0(z)$ from the leading term of that. Using the notation
   associated to (\ref{eq:Htotal}), one has
   \[ \lambda_0(z) = -\Smallfrac{1}{2} \xi^2 s_0(z) = -\Smallfrac{1}{2} \xi^2 
          \left[ \psi^2(z) + (1 -\tilde{\omega}) \psi(z) \right],\]
   \[ \mbox{where} \quad \psi(z) = \Smallfrac{1}{2} \tilde{\omega}
        \left\{1 - \frac{(\theta + \i z \rho \xi)}{\sqrt{(\theta + \i z \rho \xi)^2 +
        		 (z^2 - \i z) \xi^2} } \right\}. \]   
   Then, the prescription in the cited reference is to look for a saddle point
   $z^* = \i y^*$ in the complex $z$-plane along the purely imaginary axis (so $y^*$ is real). 
   The saddle point is a solution to $\lambda_0'(z)=0$, where the prime indicates
   the $z$-derivative. Having found that, the asymptotic $T \ra \infty$
   option implied volatility $V^{imp}_{\infty} = (\sigma^{imp}_{\infty})^2$ is given by the very simple relation
   \begin{empheq}[box =\fbox]{align}  
   	 V^{imp}_{\infty} = 8 \lambda_0(z^*). \label{eq:VimpInf} 
   	 \end{empheq}
   With $\psi^* \equiv \psi(\i y^*)$, it's easy to find 
   
   \be y^* = \frac{\theta}{2 \, \theta - \rho \xi}
    \quad \Rightarrow \quad  \psi^* = \frac{\tilde{\omega}}{2} \,
      \left\{ 1 - \zeta \right\}.
             \label{eq:deltastar} \eb  
    \[ \mbox{defining} \quad  \zeta \equiv  
    	 \frac{1}{(1+\frac{\xi^2}{4 \theta (\theta - \rho \xi)})^{1/2}}, \quad 
    	 (\theta > \rho \,\xi).     \]  
   Notice that we work in the ``martingale-preserving" regime (recall Sec. \ref{sec:martingality}),
   where $-1 \le \rho \le (\theta/\xi)$. Since we assume throughout that $\theta \ge 0$,
   in that regime $\zeta \in [0,1)$ and $y^* \in [0,1]$, which is the preferred (and
   guaranteed) analyticity strip for $H$.	 
    Thus, we have found that \emph{when $\tilde{\omega}$ is such that a
    discrete spectrum exists}, then 

 \[ \lambda_0^* \equiv \lambda_0(z^*) =  
 \Smallfrac{1}{2} \xi^2 
\left[  (\tilde{\omega} -1) \psi^* - (\psi^*)^2 \right],\]
   \[ \mbox{and} \quad  \sigma^{imp}_{\infty} =	2 \, \xi \sqrt{(\tilde{\omega}-1) \psi^* - (\psi^*)^2}, \] 
     where $\psi^*$ is given by (\ref{eq:deltastar}). But, what is the criterion for
     a discrete spectrum to exist? From (\ref{eq:Htotal}), and since $\psi^*$ is real,
     a discrete spectrum exists when
     \[     \tilde{\omega} \ge 1 + 2 \psi^* = 1 + \tilde{\omega} (1 - \zeta)
            \quad \Rightarrow \quad \tilde{\omega} \ge \frac{1}{\zeta}.  \]
    Now, since we are discussing Case 1, which has $\tilde{\omega} \ge 1$, the remaining
    sub-case to be addressed here is $1 \le \tilde{\omega} < (1/\zeta)$. Since there
    is no discrete spectrum, only the continuous spectrum
    term contributes. One reads off from  (\ref{eq:Htotal}) that
    \[ H(T,\sigma;z) \sim \e^{-(1 - \tilde{\omega})^2 \xi^2 T/8} u_0(\sigma;z); \]      
    \[ \mbox{i.e., the mass gap} \quad \lambda_0^* =  
      \Smallfrac{1}{8} \xi^2 (\tilde{\omega}-1)^2, \quad
       ( 1 \le \tilde{\omega} < \frac{1}{\zeta} ), \]  
       with some $u_0$ whose specific form doesn't matter for our purpose.
    Let's introduce the notation $\tilde{\omega}_c \equiv 1/\zeta$, which
    marks the critical value of $\tilde{\omega}$ such that a discrete spectrum emerges.
    To summarize, we've found under Case 1 (and the martingale-preserving regime) that
    the principal eigenvalue/mass gap is:  
       
    \begin{empheq}[box=\fbox]{align}   
       & \lambda^*(\tilde{\omega}) =   \Smallfrac{1}{2} \xi^2
      \left \{ 
      \begin{array}{ll}
      	 (\tilde{\omega} -1) \psi^*(\tilde{\omega}) - (\psi^*(\tilde{\omega}))^2, \quad & 
      	     \tilde{\omega}_c \le \tilde{\omega} < \infty, \\
      	 \Smallfrac{1}{4} (\tilde{\omega}-1)^2, \quad & 1 \le \tilde{\omega} < \tilde{\omega}_c,          	          
      \end{array}  \right.   \label{eq:lamstar} \\
     & \mbox{where} \,\, \tilde{\omega}_c = \frac{1}{\zeta}, \quad 
     \psi^*(\tilde{\omega}) = \frac{\tilde{\omega}}{2} \,  \left\{ 1 - \zeta \right\},
       \quad \mbox{and} \,\,\,  \zeta =   
       \frac{1}{(1+\frac{\xi^2}{4 \theta (\theta - \rho \xi)})^{1/2}}. \nonumber          
      \end{empheq}  
  
  \pbold{Smoothness.} Notice that $\psi^*(\tilde{\omega}_c) = (1-\zeta)/(2 \zeta)= 
  \Smallfrac{1}{2}(\tilde{\omega}_c -1).$ Thus
  \[    (\tilde{\omega}_c -1) \psi^*(\tilde{\omega}_c) - (\psi^*(\tilde{\omega}_c))^2   
         = \Smallfrac{1}{4} (\tilde{\omega}_c -1)^2, \]
         which shows that $\lambda^*(\tilde{\omega})$ of (\ref{eq:lamstar})
         is continuous at $\omega = \tilde{\omega}_c$. By differentiating, it is
         easy to see that $d \lambda^*(\tilde{\omega})/d \tilde{\omega}$ is
         also continuous at  $\omega = \tilde{\omega}_c$. In other 
         words $\lambda^*(\tilde{\omega})$ is \emph{smooth}, in the sense
         of being continuously differentiable, throughout the Case 1 regime: 
         $1 \le \tilde{\omega} < \infty$.

     \pbold{Examples.}
     With $\omega = \xi = 1$ and $\rho = 0$, then $\tilde{\omega}=2$,   
     $\psi^* =  0.007722$, and $\sigma^{imp}_{\infty} = 0.1751$. Or, 
       with $\omega = \xi = 1$ and $\rho = -0.75$, then   
     $\psi^* =  0.006515$, and $\sigma^{imp}_{\infty} = 0.1609$.  This is how the
     $T=\infty$ entries in Table \ref{tab:NumericalExample1} were found.

    \pbold{Case 2.}  
   For simplicity, suppose no cost-of-carry parameters, so
   $r=q=0$. As $T$ gets large under Case 2 the fundamental transform $H$ does not decay,
   but instead has a non-trivial, but relatively simply limit: 
   $H(T,\sigma_0; z) \sim H_{\infty}(z,\sigma_0)$. From (\ref{eq:putval}):   
   \be P_{\infty}(S_0,\sigma_0;K) = K \left[1 -
   \frac{1}{2 \pi} \int_\Gamma \e^{-\i z X} \frac{H_{\infty}(z,\sigma_0)}{z^2 - \i z}  \, dz \right]. \label{eq:PutFromHinf} \eb
  Here the contour $\Gamma$ is defined by $0 < \Im \, z < 1$, and $X = \log (S_0/K)$.
  Note that $H_{\infty}(z,y)$ is not a new function in our development; indeed, $H_{\infty}(z,y) = h(y;z)$, where the r.h.s. was found previously at  (\ref{eq:Hinf}). Hence,  
          
  \be  H_{\infty}(z,\sigma_0) = \e^{-\gamma_z \sigma_0}
       \times \left \{ 
          \begin{array}{ll}
              \frac{U(\psi_z,\tilde{\omega},R_z \sigma_0)}{U(\psi_z,\tilde{\omega},0)}, 
             \quad & 0 < \tilde{\omega} \le 1, \\
             1, \quad & \tilde{\omega}=0,          	          
          \end{array}  \right.  
            \label{eq:Hinf2} \eb
            and recall Sec.  \ref{sec:notations} for other notations.\footnote{The
            	case $\tilde{\omega}=0$ most easily follows by seeking a
            	stationary solution $h(y;z)$ to the top line of (\ref{eq:Greenfunc})
            	with $\omega = 0$. One easily find $h(y;z) = \e^{-\gamma_z y}$. Alternatively,
            	it can be checked by proving $\lim_{b \ra 0} U(a,b,z)/U(a,b,0) =1$.}

  Numerical examples using (\ref{eq:PutFromHinf}-\ref{eq:Hinf2}) are found in the $T=\infty$ entry in Table	\ref{tab:NumericalExample2}. A similar discussion for the SABR model
  is found in \cite{lewis:2016} (Sec. 8.12.1).          
    
    \newpage
  
  \section{Numerical examples for option prices} \label{sec:numerics}
 Formulas are implemented in Mathematica; prices use (\ref{eq:putval}). 
 All required special functions are built in.  
  We have a double integration. Additional complications that must be handled are now discussed.
  
  \pbold{Implementation notes: Case 1.}
  
  For the `physical' Case 1 ($\omega \ge \xi^2/2$), we have a double integration using confluent
  hypergeometric functions with complex arguments. Results are not immediate. 
  For example, with my choices for cutoffs and Precision parameters, option prices are found in 0.5-2 minutes on my desktop
  machine.
  
  The main complication in this case is that, prior to the Fourier integration over $z$, one must identify
  the transition points (if any: see Appendix A) where the discrete and continuous parts of $H$ are discontinuous.
  In Mathematica, I wrote a module {\texttt{GetAllTransitionPts[..]}}, which returns a list of
  such points. As I integrate in the $z$-plane over the line $z = \i/2 + x$ for $x \in [0,x_{max}]$,
  the list only includes transition points in that (finite) interval. Now, in Mathematica, when
  integrating using {\texttt{NIntegrate[..]}} and typical defaults, the integration is adaptive and you
  don't know in advance what points will be sampled. In the function returning the integrand,
   I  first determine if each sample point $x$
  is near any transition point $x^*$, nearness meaning $x \in (x^*-\epsilon,x^*+\epsilon)$, where 
   $\epsilon > 0$ is small. If not near, I just return
  the `normal' integrand. If near, I return the linearly interpolated $H$-value, using the endpoint
  values at $x_1 = x^* - \epsilon$ and $x_2 = x^* + \epsilon$. This keeps $H$ continuous, while
  avoiding troublesome evaluations of the discontinuous components too close to their
  transition points.

  \pbold{Implementation notes: Case 2.}
  
  For the unphysical Case II ($\omega < \xi^2/2$), the good news is that there are no discontinuity points. The bad news is that the integrand
  requires the infinite sum at (\ref{eq:kappadef}). I truncate that sum, and
  make some use of Mathematica's  {\texttt{Parallelize}}. This case
  is really computationally tedious: results can take a half-hour or more.
  
  \pbold{Examples.}
  
\begin{figure}[hp] 
 	   	\caption{\textbf{Large T behavior of the Implied Volatility}}
 	   	\vspace{10pt}
 	   	\includegraphics[width=0.9\textwidth]{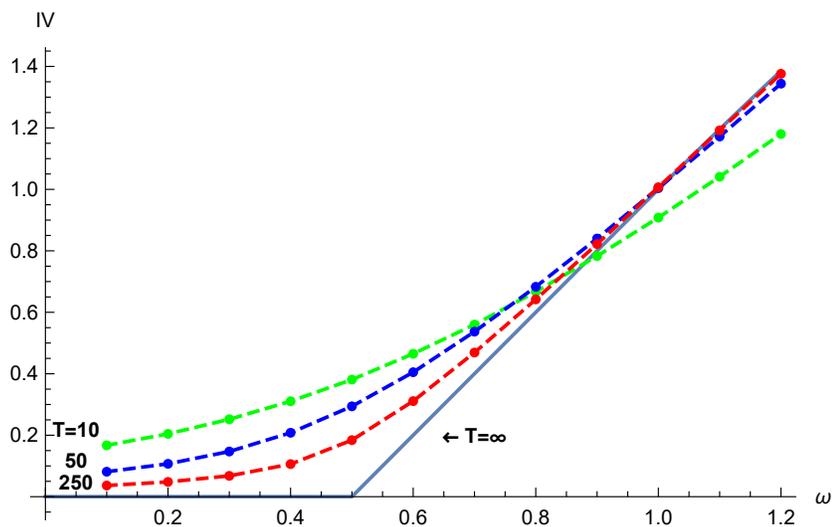}
 	   	\label{fig:XGBMLargeTIV}
 	   \end{figure}

  As a first example, see Tables \ref{tab:NumericalExample1} and \ref{tab:NumericalExample2}
  for Cases 1 and 2 respectively.  Table  \ref{tab:NumericalExample3} shows 
 additional at-the-money implied volatilities at larger $T$, the easiest regime for spectral
 solutions. There
  is good convergence to the exact analytic results at $T = \infty$, 
  found in Sec. \ref{sec:largeT}. Table \ref{tab:NumericalExample3} results are plotted Fig. \ref{fig:XGBMLargeTIV}.
   
  Various results were spot-checked for consistency with Monte Carlos. As a stronger check, Table \ref{tab:PDEcompares} shows some
  comparisons and good agreement with option values from a PDE solver.\footnote{\label{ft:PDE}The PDE solver was developed by Y. Papadopoulos, who
  provisionally extended the method in \cite{2018:paplew} to the XGBM model (private communication).}  
  
    \newpage

  \begin{table}[h] 
  	\begin{center}
  		\begin{tabular}{crlrlrl} 
  			\toprule   &   \multicolumn{2}{c}{$(\omega=1,\rho=0)$}
  				 &   \multicolumn{2}{c}{$(\omega=1,\rho=-0.75)$} & 
  	          \multicolumn{2}{c}{$(\omega=2,\rho=-0.75)$} \\
  	        \cmidrule(r){2-3}   \cmidrule(r){4-5}  \cmidrule(r){6-7} 
	                   &   Option & Implied    & Option & Implied & Option & Implied \\
	          T        &   Price  & Vol(\%)    &  Price & Vol(\%) & Price & Vol(\%) \\
	          \midrule   
	          0.25  & 4.126 &20.69  &  4.033      &  20.23         &  4.543     & 22.79   \\   
	          2     & 11.57 &20.57  & 10.78       &  19.17         &  18.15     & 32.46   \\  
	          20    & 30.97 &17.82  & 28.68       &  16.44         &  61.26     & 38.65   \\  
	          100   & 61.94 &17.54  & 58.03       &  16.14         &  95.06     & 39.30   \\  
	          500   & 94.98 &17.51  & 92.81       &  16.10         &  99.999    & 39.42   \\  
	        $\infty$& 100   & 17.51 & 100       &  16.09         &   100      & 39.46   \\  
  					\bottomrule 
  		\end{tabular}
  	\end{center}
  	\caption{{\bf{At-the-money option values for Case 1: $\mathbf{\omega \ge \xi^2/2}$.}} 
  		Other risk-neutral model parameters: $S_0=K=100$, $\sigma_0 = 0.20$, $\xi=1$,
  		 $\theta=4$,  $r=q=0$. Under the stationary density, $\Ex{\sigma}=(0.125,0.375)$ for $\omega=(1,2)$  respectively. The $T=\infty$ Implied Vol entries are found from the square-root of (\ref{eq:VimpInf}).} 
  	\label{tab:NumericalExample1}
  \end{table}

   \begin{table}[h] 
  	\begin{center}
  		\begin{tabular}{crlrlrl} 
  			\toprule  	 &   \multicolumn{2}{c}{$(\omega=\Smallfrac{1}{4},\theta=4)$} & 
  			\multicolumn{2}{c}{$(\omega=\Smallfrac{1}{2},\theta=4)$} &
  			\multicolumn{2}{c}{$(\omega=0, \theta=0)$} \\
  			\cmidrule(r){2-3}   \cmidrule(r){4-5}  \cmidrule(r){6-7}
  			       &   Option  & Implied   
  			 & Option  & Implied   & Option & Implied \\
  			   T        &   Price  & Vol(\%)    &  Price & Vol(\%) & Price & Vol(\%) \\
  			\midrule   
  			0.25     & 3.705  & 18.58  & 3.809 & 19.11   & 3.962 & 19.87             \\   
  			2        & 7.287  & 12.93  & 8.268 & 14.68   & 10.35      & 18.39             \\  
  			20       & 9.577  & 5.38   & 13.56 & 7.64   & 15.49      & 8.74             \\  
  			100      & 9.756  & 2.45   & 18.50 & 4.68  & 15.554      & 3.92            \\  
  			500      & 9.758  & 1.10   & 25.82 & 2.95   & 15.554     & 1.75             \\  
  			$\infty$ & 9.758  & 0      & 100   & 0        & 15.554     & 0            \\  
  			\bottomrule 
  		\end{tabular}
  	\end{center}
  	\caption{{\bf{At-the-money option values for Case 2: $\mathbf{\omega < \xi^2/2}$.}} 
  		Other risk-neutral model parameters: $S_0=K=100$, $\sigma_0 = 0.20$, $\xi=1$, $\rho=-0.75$,  $r=q=0$.
  		The middle column is technically a Case 1 edge case, but it was computed two ways
  		with the same result: (i) using Case 1 code with $\omega = 0.5$ and (ii) using
  		Case 2 code with $\omega = 0.49999$. For (ii), the sums in (\ref{eq:kappadef})
  		were truncated at $n=10$ terms. The $T=\infty$ price entries are computed from (\ref{eq:PutFromHinf}).} 
  	\label{tab:NumericalExample2}
  \end{table}
  
 \newpage
 
  \begin{table}[h] 
 	\begin{center}
 		\begin{tabular}{rcccccccccccc} 
 				\multicolumn{13}{c}{Implied Volatility (decimal)} \\
 			\toprule  
 			& 	\multicolumn{4}{c}{Case 2} &  	\multicolumn{8}{c}{Case 1} \\
 			 \cmidrule(r){2-5}   \cmidrule(r){6-13}  
 			     T &$\omega=0.1$ & 0.2 & 0.3 & 0.4 & 0.5 & 0.6 & 0.7 & 0.8 & 0.9 & 1.0 &1.1 &1.2 \\
 			    \cmidrule(r){2-5}   \cmidrule(r){6-13}
  $10$& 0.1672 & 0.2044  & 0.2518 & 0.3105  & 0.381  & 0.465& 0.560  & 0.667& 0.783 &0.908 &1.041 &1.180 \\ 			  
  $50$ & 0.0813  & 0.1070  & 0.1470 & 0.2080  & 0.294  & 0.405& 0.537  & 0.683& 0.840 &1.004&1.172 &1.344 \\
  $250$& 0.0364  &0.0481   & 0.0678 & 0.1061  & 0.184  & 0.311& 0.469  & 0.642 &0.822 &1.007&1.192&1.376\\
 		      $\infty$ &0  &0  &0  &0  &0  &0.2  &0.4  &0.6  &0.8  &1.0 &1.196 &1.386  \\
 			\bottomrule 
 		\end{tabular}
 	\end{center}
 	\caption{{\bf{Large $T$ behavior of the Implied Volatility.}} 
 		Other risk-neutral model parameters: $S_0=K=100$, $\sigma_0 = 0.20$, 
 		  $\theta=1/\sqrt{12}$, $\xi=1$, $\rho=0$,  $r=q=0$. The $T=\infty$ entries 
 		  are computed from the exact relations in Sec. \ref{sec:largeT}. The
 		  data are plotted in Fig. \ref{fig:XGBMLargeTIV}.} 
 	\label{tab:NumericalExample3}
 \end{table}

 \newpage
 
  \begin{table}[h] 
 	\begin{center}
 		\begin{tabular}{crlrl} 
 			\multicolumn{5}{c}{Implied Volatility at Various Strikes} \\
 			\toprule   &   \multicolumn{2}{c}{Case A}
 			&   \multicolumn{2}{c}{Case B} \\			
 			\cmidrule(r){2-3}   \cmidrule(r){4-5}  
 			Strike        &   PDE  & Exact    &  PDE & Exact \\
 		    \cmidrule(r){1-1}	\cmidrule(r){2-3}   \cmidrule(r){4-5}    
 			3400   & 0.6384 & 0.6384  &       &            \\   
 			3600   & 0.5854 & 0.5854  & 0.2813  &  0.2812        \\  
 			3800   & 0.5322 & 0.5322  & 0.2765  &  0.2765       \\  
 			4000   & 0.4787 & 0.4787  & 0.2719  &  0.2719       \\  
 			4200   & 0.4252 & 0.4252  & 0.2675  &  0.2676       \\  
 			4400   & 0.3742 & 0.3742  & 0.2634  &  0.2635          \\  
 			4600   & 0.3344 & 0.3344  & 0.2594  &  0.2595         \\  
 			4800   & 0.3196 & 0.3196  & 0.2557  &  0.2558           \\  
 			5000   & 0.3263 & 0.3263  & 0.2522  &  0.2524            \\  
 			5200   & 0.3420 & 0.3420  & 0.2490  &  0.2492         \\  
 			5400   & 0.3604 & 0.3604  & 0.2462  &  0.2463            \\  
 			5600   & 0.3792 & 0.3792  & 0.2437  &  0.2437          \\  
 			\bottomrule 
 		\end{tabular}
 	\end{center}
 	\caption{{\bf{Numerical check: PDE vs. (quasi) exact solutions.}} Table entries show
 		the implied volatility (decimal) for two cases.	For Case A: $T=\frac{14}{365}, \,\, r = 0.0357$. For Case B: $T=\frac{714}{365}, \,\,
 		 r = 0.0401$. 
 		Common parameters: $S_0=4468.17$, $q=0$, $\sigma_0 = 0.365113$, 
 		$\omega=24.8424$, $\theta=63.2858$, $\rho=-0.517545$, $\xi=5.33384$. As $2 \omega/\xi^2 > 1$, 
 		this is a Case 1 comparison. Parameters and PDE values come from an XGBM model calibration against a (July 5, 2002) DAX index option chain using software developed by Y. Papadopoulos (footnote \ref{ft:PDE}). Exact values are implied vols using prices computed from (\ref{eq:putval}).} 
 	\label{tab:PDEcompares}
 \end{table}

 \section{Other applications: $P$-model parameter estimation} 
   Maximum likelihood estimation (MLE) is the preferred approach -- feasible using the transition density at (\ref{eq:solnfinal}) with a volatility proxy for the $y$'s. 
  Key are dimensionless ratios such as
    $\sigma_i^2 (\theta/\xi)^2 (T_i-T_{i-1})$  and  $\sigma_i (\omega \theta/\xi^2) (T_i-T_{i-1})$,
    where $\sigma_i = \sigma(T_i)$. 
     If the data observations are such that these ratios
     are not particularly small, then a fast and efficient (say C/C++) implementation 
  is likely needed. However, when the $\mbox{ratios} \ll 1$,
 as may be the case with daily or weekly observations, then small-time asymptotics
 for the transition density should be effective and the exact formulas can simply serve
 as checks. Results along these lines will be reported in another publication.

\newpage

% \bibliographystyle{plain}   
% \bibliography{alantexstudio}   

 \newpage
 \section{Appendix A -- Proof of Theorem 1}
  The goal is to solve problem (\ref{eq:Greenfunc}) for $G(\tau,y,y')$. With $\Re s > 0$, the Laplace transform
   $\hat{G}(s,y,y') = \int \e^{-s \tau} G(\tau,y,y') \, d \tau$ exists and solves
  \be  y^2 \hat{G}_{yy} + (\omega y - \theta y^2) \hat{G}_y - (s +c y^2) \hat{G} = -\delta(y-y'),
       \quad (y \in \R_+). 
         \label{eq:Gode} \eb
   The associated homogeneous ODE 
   \be y^2 u'' + (\omega y - \theta y^2) u' - (s +c y^2) u = 0  \label{eq:homoode} \eb
   is converted
   to Kummer's differential equation in two steps. First, suppressing the dependence on $(s,y')$, 
   let $u(y) = \e^{-\gamma y} y^r f(y)$, choosing
   \be \gamma = -\Smallfrac{1}{2} (\theta - \sqrt{\theta^2 + 4 c}) \quad \mbox{and} \quad 
        r = r_s = \Smallfrac{1}{2} (1 - \omega) +  \Smallfrac{1}{2} \sqrt{(1-\omega)^2 + 4 s}. 
          \label{eq:appAparmdefs1} \eb
    This yields the ODE
    \[  y \, f'' + [ (2 r + \omega) - (2 \gamma + \theta) y ] f' - [ r (2 \gamma + \theta) + \omega \gamma] f=0. \]
    Second, let $f(y) = w(x)$, where $x = R y$, choosing 
    \be R = \theta + 2 \gamma = \sqrt{\theta^2 + 4 c}. \label{eq:appAparmdefs2} \eb
    With that,
    $w(x)$ solves Kummer's differential equation
    \[ x w_{xx} + (b - x) w_x - a w =0, \,\, \mbox{with} \,\, b = b_s = \omega + 2 r, \,\,
     \mbox{and} \,\,      a = a_s = r + \frac{\omega \gamma}{R}.  \]
    Note that in the case of real coefficients $(\theta,c)$ with $c > 0$, we have
    both $\gamma > 0$ and $R > 0$, regardless of the sign of $\theta$. To
    construct $\hat{G}$, we need solutions to Kummer's equation that are relatively small
    as $x \ra 0$ and $x \ra +\infty$ respectively. Suitable choices here are the confluent hypergeometric
    functions $M(a,b,x)$ and $U(a,b,x)$. We denote the associated solutions to (\ref{eq:homoode})
    as $\xi(y,s)$ and $\eta(y,s)$. Using the above notations,
    \be  \xi(y,s) = \e^{-\gamma y} ( R y)^r M(a,b,R y) \quad \mbox{and} \quad  
      \eta(y,s) = \e^{-\gamma y} ( R y)^r \, U(a,b,R y). \label{eq:appAsolndefs} \eb  
    In terms of those, the solution to (\ref{eq:Gode}) has the well-known form:
    \be \hat{G}(s,y,y') =  \frac{-1}{(y')^2 W(y',s)} \left \{
    \begin{array}{l}
    	\xi(y,s) \, \eta(y',s), \quad 0 \le y \le y',   \\
    	\xi(y',s) \, \eta(y,s),  \quad y' \le y \le \infty,     	
    \end{array}  \right.  \eb
where $W(y,s) \equiv \xi \, \eta_y - \eta \, \xi_y$ is the Wronskian of $(\xi,\eta)$. With that, a first solution
for the Green function $G(\tau,y,y')$ is given by the Bromwich Laplace inversion
\be  G(\tau,y,y')  =   \int_{\Gamma} \frac{ds}{2 \pi \i} \, \e^{s \tau} \, \hat{G}(s,y,y'), \eb
where  $\Gamma$ is a vertical line in the complex $s$-plane to the right of any singularities.
As $G(\tau,y,y')$ is a probability transition density, bounded on $\tau > 0$,
then $\hat{G}(s,y,y') = \int \e^{-s \tau} G(\tau,y,y') \, d \tau$ is regular for all $\Re s > 0$. Thus, any vertical contour to the right of $s=0$ suffices
for $\Gamma$: see Fig. \ref{fig:XGBMInversionContour}. 

\newpage
\Pbreak
Using Wronskians found in \cite{as:1972}, we find

\be W(y,s) = -\frac{\Gamma(b_s)}{\Gamma(a_s)} R (R y)^{-\omega} \e^{\theta y}   
         \label{eq:wronskians} \eb
          Summarizing at this point, we have

\Pbreak
\Pbreak
{\underline{First solution (Auxiliary Green function by Bromwich inversion):}} 
\begin{empheq}[box=\fbox]{align} 
	G(\tau,y,y')  =&   \int_{\Gamma} \frac{ds}{2 \pi \i}  \, \e^{s \tau} \, \hat{G}(s,y,y'),
	       \,\, \mbox{where} \,\, \Gamma \,\, \mbox{is in Fig.} \ref{fig:XGBMInversionContour}
	       \,\, \mbox{and} \nonumber \\ 
	      \nonumber \\
	 \hat{G}(s,y,y') &=  \frac{\Gamma(a_s)}{\Gamma(b_s)} R^{b_s-1} y^{r_s} (y')^{r_s + \omega-2}
	  \e^{-\gamma y} \e^{-(\gamma + \theta) y'}  \nonumber \\
  &\times \left \{
\begin{array}{l}
	M(a_s,b_s,R y) \, U(a_s,b_s, R y'), \quad 0 \le y \le y',   \\
	M(a_s,b_s, R y') \, U(a_s,b_s, R y),   \quad y' \le y \le \infty,     	
\end{array}  \right. \label{eq:hatGfirstsol}  \\
    \nonumber \\
  \mbox{using} \quad R &=   \sqrt{\theta^2 + 4 c}, \quad 
 \gamma = -\Smallfrac{1}{2} (\theta - \sqrt{\theta^2 + 4 c}), \nonumber \\
  b_s &= 1 + \sqrt{(1-\omega)^2 + 4 s}, \quad a_s = r_s + \frac{\omega \gamma}{R}, \nonumber \\
  \mbox{and} \quad r_s &= \Smallfrac{1}{2} (1 - \omega) +  \Smallfrac{1}{2} \sqrt{(1-\omega)^2 + 4 s}. \nonumber
 \end{empheq}
  
\subsection{Spectral solution, Case 2}
\begin{figure}[hp] 
 		   	\caption{\bf{Green function inversion contour in the complex $s$-plane}}
 		   	\vspace{10pt}
 		   	\begin{center}
 		   		\includegraphics[width=0.8\textwidth]{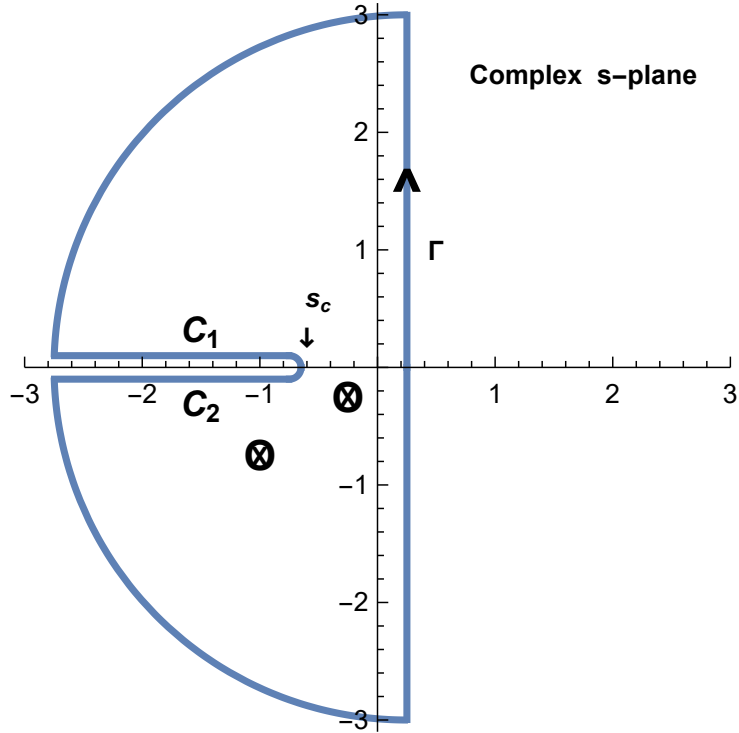}
 		   		\newline\newline
 		   	\end{center}		
 		   	\label{fig:XGBMInversionContour}
 		   	{\bf{Notes.}} The vertical contour $\Gamma$ is the Bromwich (Laplace) inversion contour in the complex $s$-plane for the auxiliary Green function $G$. The Residue Theorem converts the integral along $\Gamma$ into an integral along the branch-cut (which extends along the line $s \in (-\infty,s_c)$ plus contributions from poles (shown as crossed-circles). The figure is schematic: there generally may be zero or more poles. The branch-cut endpoint is at $s_c = -\frac{1}{4}(1-\tilde{\omega})^2 \le 0$, where $\tilde{\omega} = 2 \omega/\xi^2$. The poles move in the $s$-plane as the separate Fourier inversion parameter $z$ moves in the complex $z$-plane. However, as $z$ changes (while other parameters are fixed), the branch-cut endpoint $s_c$ remains fixed.
 		   \end{figure}

Case 2 is simpler in some respects, so a good place to begin. 
We construct the spectral solution from (\ref{eq:hatGfirstsol}) via the Residue Theorem, extending the
Bromwich contour to the closed path shown in Fig. \ref{fig:XGBMInversionContour}. The precise nature
of that path is explained below.

As one sees from
 (\ref{eq:hatGfirstsol}), $\hat{G}(s,y,y')$ is dependent upon the parameters $b_s$ and $r_s$. Both
 have a (square-root) branch point singularity at $s = s_c \equiv -\frac{1}{4}(1-\omega)^2 \le 0$,
 using `c' for `cut'. 
 Regardless of the value of (real) $\omega$, this branch point is always present in the
 left half $s$-plane on the negative axis. We construct a branch cut along the negative $s$ axis, extending from  $s_c$ to $s = -\infty$. Then, we integrate infinitesimally above and below the cut, along
 the contours $C_1$ and $C_2$ (Fig. \ref{fig:XGBMInversionContour}).    
 
 Now $\hat{G}$ may also have pole singularities \emph{inside} the closed contour.
 Note the branch-cut is \emph{outside}. However, it will be shown below that there
 are no such poles when $0 \le \omega < 1$, our Case 2. It can also be shown
 that the contribution from the integrations along the curved quarter-circles of
 Fig. \ref{fig:XGBMInversionContour} vanish as these contour segments  extend to $|s| \ra \infty$.
 Hence, in that case,  $\hat{G}(s,y,y')$ is regular everywhere inside the closed contour of the figure: 
 we have for Case 2, by the Residue Theorem,
 
\be G(\tau,y,y')  =   \int_{\Gamma} \frac{ds}{2 \pi \i}  \, \e^{s \tau} \, \hat{G}(s,y,y')
   = -  \int_{C_1 + C_2} \frac{ds}{2 \pi \i}  \, \e^{s \tau} \, \hat{G}(s,y,y'). \label{eq:ResThrm} \eb
 Now change integration variables from $s$ to $\nu$ by letting
 $s = -\frac{1}{4}(1-\omega)^2 - \nu^2 \pm \i \epsilon$ along $C_1$ and $C_2$ respectfully.
 Then, as $\epsilon \downarrow 0$, 
 
 \[ \mbox{along} \,\, C_1: \quad r_s \ra r^+_{\nu} \equiv \Smallfrac{1}{2} (1 - \omega) + \i \nu,
       \quad b_s \ra b^+_{\nu} \equiv 1 + 2 \i \nu, \quad a_s \ra a^+_{\nu} \equiv r^+_{\nu} + \psi_z, \]
       abbreviating $\psi_z =  \omega \gamma_z/R_z$. Importantly, while $\psi_z$ generally
       depends upon the Fourier parameter $z$, it is \emph{independent} of the Laplace parameter $s$.        
     Similarly, as $\epsilon \downarrow 0$, 
      \[ \mbox{along} \,\, C_2: \quad r_s \ra r^-_{\nu} \equiv \Smallfrac{1}{2} (1 - \omega) - \i \nu,
       \quad b_s \ra b^-_{\nu} \equiv 1 - 2 \i \nu, \quad a_s \ra a^-_{\nu} \equiv r^-_{\nu} + \psi_z. \]
    Notice that, since  $ a^{\pm}_{\nu} = \Smallfrac{1}{2} (1 - \omega) \pm \i \nu + \psi_z$, we have
    for every fixed $z$,
    \be  a^-_{\nu} + 2 \i \nu = a^+_{\nu}. \label{eq:keyparmrelation} \eb   
   Suppressing dependencies on $z$, recall that   
     \[  \eta(y,s) = \e^{-\gamma y} ( R y)^{r_s} U(a_s,b_s, R y).   \]
     Using the  representation (\ref{eq:Udef}) and abbreviating $\zeta \equiv R y$, along $C_1$:
     \bea && \eta(y,s) \ra \eta^{+}(y,\nu) \equiv
          \e^{-\gamma y} \zeta^{r^+_{\nu}} U(a^+_{\nu},b^+_{\nu},\zeta) \label{eq:etarep1} \\
         &=&  \e^{-\gamma y} \zeta^{(1 - \omega)/2 + \i \nu} \,
         U(a^+_{\nu}, 1 + 2 \i \nu,\zeta) \nonumber \\
         &=& \e^{-\gamma y} \zeta^{(1 - \omega)/2} \nonumber \\
      &\times&  \left[ \zeta^{\i \nu} \frac{\Gamma(-2 \i \nu)}{\Gamma(a^+_{\nu}-2 \i \nu)}
       M(a^+_{\nu}, 1 + 2 \i \nu, \zeta) 
      + \zeta^{-\i \nu} \frac{\Gamma(2 \i \nu)}{\Gamma(a^+_{\nu})} 
      M(a^+_{\nu}-2 \i \nu, 1 - 2 \i \nu, \zeta) \right].
         \nonumber \aeb 
   Similarly, along $C_2$:
         \bea && \eta(y,s) \ra \eta^{-}(y,\nu) \equiv
         \e^{-\gamma y} \zeta^{r^-_{\nu}} U(a^-_{\nu},b^-_{\nu},\zeta) \nonumber \\
         &=&  \e^{-\gamma y} \zeta^{(1 - \omega)/2 - \i \nu} \,
         U(a^-_{\nu}, 1 - 2 \i \nu,\zeta) \nonumber \\
         &=& \e^{-\gamma y} \zeta^{(1 - \omega)/2} \nonumber \\
         &\times&  \left[ \zeta^{-\i \nu} \frac{\Gamma(2 \i \nu)}{\Gamma(a^-_{\nu}+2 \i \nu)}
         M(a^-_{\nu}, 1 - 2 \i \nu, \zeta) 
         + \zeta^{\i \nu} \frac{\Gamma(-2 \i \nu)}{\Gamma(a^-_{\nu})} 
         M(a^-_{\nu}+2 \i \nu, 1 + 2 \i \nu, \zeta) \right].
         \nonumber \aeb 
  Comparing, in light of (\ref{eq:keyparmrelation}), we have the key relation
  \begin{empheq}[box=\fbox]{align}
      \eta^{+}(y,\nu) =   \eta^{-}(y,\nu).   
      \end{empheq}
  In other words, $\eta(y,\cdot)$ is the same function of dummy integration variable $\nu$ above and below the cut.\footnote{Another thing to notice is that, when the parameters $(\omega,\theta,c)$ are \emph{real},
  	then $\eta(y,s)$ is also real along the cut. This follows because, under those assumptions,
  	we have $(a^+_{\nu})^* =  a^-_{\nu}$ and $(b^+_{\nu})^* =  b^-_{\nu}$, where the asterisk denotes
  	complex conjugation. However,  $(a^+_{\nu})^* \not=  a^-_{\nu}$ generally in our application. To
  	elaborate, that's because we have  $(\omega,\theta,c) \ra (\omega,\theta_z,c_z)$ and,
  	in general (with non-zero correlation) $\theta_z = \Smallfrac{2}{\xi^2}  (\theta + \i z \rho \xi),$ is not real along any Fourier inversion contour in the complex $z$-plane, although $\theta$ itself is real. In turn, this causes $\psi_z$ to not be real, which causes 
  	 $(a^+_{\nu})^* \not=  a^-_{\nu}$.}  To see the implications of that, take the case where $y \le y'$.
   Then, we have
   
   \begin{empheq}{align*}
   G(\tau,y,y')	&= -  \int_{C_1 + C_2}  \e^{s \tau} \, \hat{G}(s,y,y') \frac{ds}{2 \pi \i}  \\      
   &= (y')^{-2} \e^{-(1-\omega)^2 \tau/4} \int_0^{\infty}  \e^{-\nu^2 \tau} \eta^+(y',\nu) 
   \left[ \frac{\xi^+(y,\nu)}{W^+(y',\nu)} - \frac{\xi^-(y,\nu)}{W^-(y',\nu)} \right]
       \frac{\nu d \nu}{\pi \i},          
   \end{empheq} 
where 
\begin{empheq}{align*}
 \xi^{+}(y,\nu) &\equiv
  \e^{-\gamma y} \zeta^{r^+_{\nu}} M(a^+_{\nu},b^+_{\nu},\zeta) \\
   &= \e^{-\gamma y} \zeta^{(1 - \omega)/2} 
    \zeta^{\i \nu}  M(a^+_{\nu}, 1 + 2 \i \nu, \zeta), 
    \end{empheq}
and
   \begin{empheq}{align*}
   	     	\xi^{-}(y,\nu) &\equiv
      	\e^{-\gamma y} \zeta^{r^-_{\nu}} M(a^-_{\nu},b^-_{\nu},\zeta) \\
      	&= \e^{-\gamma y} \zeta^{(1 - \omega)/2} 
      	\zeta^{-\i \nu}  M(a^-_{\nu}, 1 - 2 \i \nu, \zeta). 
      \end{empheq}  
  And from (\ref{eq:wronskians}),  
 \[  W^{\pm}(y,\nu) = -R (R y)^{-\omega} \e^{\theta y} 
       \frac{\Gamma(b^{\pm}_{\nu})}{\Gamma(a^{\pm}_{\nu})}. \]
 Thus,
  \begin{empheq}{align}
 	&G(\tau,y,y')      
 	= -(y')^{-2} {R}^{-1} (R y')^{\omega} \e^{-\theta y'}  \e^{-\gamma y} \zeta^{(1 - \omega)/2} \e^{-(1-\omega)^2 \tau/4} \nonumber \\
 	&\times \int_0^{\infty}  \e^{-\nu^2 \tau} \eta^+(y',\nu) 
 	\left[ \frac{\Gamma(a^+_{\nu})}{\Gamma(b^+_{\nu})} \zeta^{\i \nu}  M(a^+_{\nu}, 1 + 2 \i \nu, \zeta) -  \frac{\Gamma(a^-_{\nu})}{\Gamma(b^-_{\nu})} \zeta^{-\i \nu}  M(a^-_{\nu}, 1 - 2 \i \nu, \zeta) \right]
 	\frac{\nu d \nu}{\pi \i},  \label{eq:Grep1}        
 \end{empheq}
But
 \begin{empheq}{align*}
 &	 \frac{\Gamma(a^+_{\nu})}{\Gamma(b^+_{\nu})} \zeta^{\i \nu}  M(a^+_{\nu}, 1 + 2 \i \nu, \zeta) -  \frac{\Gamma(a^-_{\nu})}{\Gamma(b^-_{\nu})} \zeta^{-\i \nu}  M(a^-_{\nu}, 1 - 2 \i \nu, \zeta)     \\
 &=   \frac{\Gamma(a^+_{\nu})}{\Gamma(b^+_{\nu})} \zeta^{\i \nu} 
      \left[ M(a^+_{\nu}, 1 + 2 \i \nu, \zeta) -  \frac{\Gamma(b^+_{\nu}) \Gamma(a^-_{\nu})}{\Gamma(a^+_{\nu}) \Gamma(b^-_{\nu})}
       \zeta^{-2 \i \nu}  M(a^-_{\nu}, 1 - 2 \i \nu, \zeta) \right]  \\
      &=   \frac{\Gamma(a^+_{\nu})}{\Gamma(b^+_{\nu})} \zeta^{\i \nu} 
      \left[ M(a^+_{\nu}, b^+_{\nu}, \zeta) -  \frac{\Gamma(b^+_{\nu}) \Gamma(a^-_{\nu})}{\Gamma(a^+_{\nu}) \Gamma(b^-_{\nu})}
      \zeta^{1- b^+_{\nu}}  M(a^+_{\nu} - b^+_{\nu}+1, 2 - b^+_{\nu}, \zeta) \right]  
 \end{empheq}
From the representation (\ref{eq:Udef}), 
\be U(a,b,z) = \frac{\Gamma(1-b)}{\Gamma(a-b+1)} \left[ M(a,b,z) 
+ z^{1-b}  \frac{\Gamma(b-1) \Gamma(a-b+1)}{\Gamma(a) \Gamma(1-b)} M(a-b+1,2-b,z) \right]. \label{eq:Udef2} \eb  
But
\[  \frac{\Gamma(b^+_{\nu})}{\Gamma(b^-_{\nu})} = -\frac{\Gamma(2 \i \nu)}{\Gamma(-2 \i \nu)} =
          -\frac{\Gamma(b^+_{\nu}-1)}{\Gamma(1-b^+_{\nu})} \]  
 Thus,         
       \begin{empheq}{align*}
            &  \frac{\Gamma(a^+_{\nu})}{\Gamma(b^+_{\nu})} \zeta^{\i \nu} 
       		\left[ M(a^+_{\nu}, b^+_{\nu}, \zeta) -  \frac{\Gamma(b^+_{\nu}) \Gamma(a^-_{\nu})}{\Gamma(a^+_{\nu}) \Gamma(b^-_{\nu})}
       		\zeta^{1- b^+_{\nu}}  M(a^+_{\nu} - b^+_{\nu}+1, 2 - b^+_{\nu}, \zeta) \right] \\ 
       		&= \frac{\Gamma(a^+_{\nu})}{\Gamma(b^+_{\nu})} \zeta^{\i \nu} 
       		\left[ M(a^+_{\nu}, b^+_{\nu}, \zeta) +	\zeta^{1- b^+_{\nu}}  \frac{\Gamma(b^+_{\nu}-1) \Gamma(a^-_{\nu})}{\Gamma(a^+_{\nu}) \Gamma(1-b^+_{\nu})}
       	  M(a^+_{\nu} - b^+_{\nu}+1, 2 - b^+_{\nu}, \zeta) \right] \\
       	  	&= \zeta^{\i \nu} \frac{\Gamma(a^+_{\nu})}{\Gamma(b^+_{\nu})}
       	  \left[ M(a^+_{\nu}, b^+_{\nu}, \zeta) +	\zeta^{1- b^+_{\nu}}  \frac{\Gamma(b^+_{\nu}-1) \Gamma(a^+_{\nu}-b^+_{\nu}-1)}{\Gamma(a^+_{\nu}) \Gamma(1-b^+_{\nu})}
       	  M(a^+_{\nu} - b^+_{\nu}+1, 2 - b^+_{\nu}, \zeta) \right] \\
       	  	&= \zeta^{\i \nu} \frac{\Gamma(a^+_{\nu}) \Gamma(a^-_{\nu})}{\Gamma(b^+_{\nu}) \Gamma(1-b^+_{\nu})}
       	  U(a^+_{\nu}, b^+_{\nu}, \zeta),
       	\end{empheq}
  where the last equality follows from (\ref{eq:Udef2}). Using this last equality and (\ref{eq:etarep1}), now (\ref{eq:Grep1}) reads 
  
  \begin{empheq}{align}
  	&G(\tau,y,y')      
  	= (y')^{-2} {R}^{-1} (R y')^{(1+\omega)/2} \e^{-\theta y'}  \e^{-\gamma (y+y')} \zeta^{(1 - \omega)/2} \e^{-(1-\omega)^2 \tau/4} \nonumber \\
  	&\times \int_0^{\infty}  \e^{-\nu^2 \tau}  
   \frac{\Gamma(a^+_{\nu}) \Gamma(a^-_{\nu})}{\Gamma(2 \i \nu) \Gamma(-2 \i \nu)} \,\,
   (\zeta \zeta')^{\i \nu} \, U(a^+_{\nu}, b^+_{\nu}, \zeta) \, U(a^+_{\nu}, b^+_{\nu}, \zeta')
  \,	\frac{d \nu}{2 \pi}, \nonumber \\  
       &= (y')^{\omega-2} (y y')^{(1-\omega)/2}  \e^{-\gamma y} \e^{-(\theta + \gamma) y'}   \e^{-(1-\omega)^2 \tau/4} \nonumber \\
        &\times \int_0^{\infty}  \e^{-\nu^2 \tau}  
        \frac{\Gamma(a^+_{\nu}) \Gamma(a^-_{\nu})}{|\Gamma(2 \i \nu)|^2}
       \,\, (\zeta \zeta')^{\i \nu} \, U(a^+_{\nu}, b^+_{\nu}, \zeta) \, U(a^+_{\nu}, b^+_{\nu}, \zeta')
        \,	\frac{d \nu}{2 \pi},  \label{eq:Grep2} 
  \end{empheq}
The relation (\ref{eq:Grep2}) gives the continuous spectrum contribution to $G(\tau,y,y')$.
The computation has been done assuming $y \le y'$, but the reader can repeat it for $y \ge y'$
and find  (\ref{eq:Grep2}) again. Also, remember that we treat two cases: Case 1, where $1 \le \omega < \infty$ and Case 2, where $0 \le \omega <1$.
We will establish below that, for Case 2, the continuous spectrum contribution is the sole
contribution. Accepting that, we can summarize at this point with

\Pbreak
{\underline{Spectral solution, Case 2 ($0 \le \omega <1$):}} 
\begin{empheq}[box=\widefbox]{align} 
	G_c(\tau,y,y')  =& (y')^{\omega-2} (y y')^{(1-\omega)/2}  \e^{-\gamma y} \e^{-(\theta + \gamma) y'}   \e^{-(1-\omega)^2 \tau/4} \int_0^{\infty} \phi_{\nu}(y,y')  \, \e^{-\nu^2 \tau}  
	\,	d \nu, \label{eq:GspectralCase2}    \\
	\mbox{using} \quad R &=  \Smallfrac{1}{2} \sqrt{\theta^2 + 4 c}, \quad 
	\gamma = -\Smallfrac{1}{2} (\theta - \sqrt{\theta^2 + 4 c}), \quad \psi =  \frac{\omega \gamma}{R},
	\nonumber \\
	a_{\nu} &= \Smallfrac{1}{2} (1 - \omega) + \psi + \i \nu, \quad b_{\nu} = 1 + 2 \i \nu, \nonumber \\
	\mbox{and} \quad \phi_{\nu}(y,y') &= \frac{1}{2 \pi}  \frac{\Gamma(a_{\nu}) \Gamma(a_{-\nu})}{\left| \Gamma(2 i \nu) \right|^2} \,
	(R^2 y y')^{\i \nu} \, U \left( a_{\nu}, b_{\nu}, R y \right) \, U \left( a_{\nu}, b_{\nu}, R y' \right). \nonumber	
\end{empheq}

\pbold{Remarks.} 

(i) The subscript `c' in (\ref{eq:GspectralCase2}) refers to the `continuous'
spectrum, which is the entire spectrum under Case 2. Later, we use `d' for the `discrete' spectrum
contribution.  

(ii) Recall that the speed density $m(y)$ associated to a diffusion is
proportional to the stationary density, given for our problem at (\ref{eq:stationayvol}).
If you remove $m(y') \propto  (y')^{\omega-2} \e^{-\theta y'}$
from (\ref{eq:GspectralCase2}), what remains is invariant under $y \leftrightarrow y'$,
a well-known general property of diffusions.

\subsection{Spectral solution, Case 1}
Referring to Fig. \ref{fig:XGBMInversionContour}, let $\mathcal{C}$ denote the entire (counter-clockwise)
closed contour and $\Omega$ the open interior bounded by $\mathcal{C}$.
In addition to the identified branch cut singularity at $s = s_c$, $\hat{G}(s,y,y')$ may have
simple pole singularities at various $s_n \in \Omega$. In that case, applying the Residue
Theorem generalizes (\ref{eq:ResThrm}) to:

\be G(\tau,y,y')  =   \int_{\Gamma} \frac{ds}{2 \pi \i}  \, \e^{s \tau} \, \hat{G}(s,y,y')
= G_c(\tau,y,y') + \sum_{s_n \in \Omega} \e^{s_n \tau} 
       \mbox{Res}[\hat{G}(s_n,y,y'] \label{eq:ResThrm2}. \eb
       Here
       \[ \mbox{Res}[\hat{G}(s_n,y,y'] 
              \equiv \lim_{s \ra s_n} (s - s_n) \, \hat{G}(s,y,y') \]
 denotes the simple pole residue of $\hat{G}$ at $s_n$ and $G_c$ is already given at              
 (\ref{eq:GspectralCase2}). The next step is locating these putative poles.
 
 Looking at (\ref{eq:hatGfirstsol}), there are potentially poles lurking in $M(a_s,b_s,\zeta)$
 and $\Gamma(a_s)$. However, while $M(a,b,z)$ has a simple pole at $b = -n$, where $n$ is a positive integer
 (see \cite{as:1972}) the function actually appears in the regularized form $M(a,b,\zeta)/\Gamma(b)$, which
 is everywhere regular. This leads us to seek possible poles in $\Gamma(a_s)$, which has simple
 poles should $a_s = -n$, where $n = 0, 1, 2, \cdots$, here a generally terminating sequence. These
 poles correspond to the vanishing of the
 Wronskian in (\ref{eq:wronskians}). More specifically, we have the  
 
 \be \mbox{\underline{pole criteria}:} \,\, s = s_n, \,\, \mbox{such that} \,\, a(s_n) = -n, \,\, 
 (n=0,1,2, \cdots) \,\,  \mbox{and} \,\, s \in \Omega. \label{eq:polecriteria} \eb 
 As we shall see, assuming $1 \le \omega < \infty$ (Case 1) and dependent upon the other parameters, there may exist zero or a finite number of such $s_n$.
 
 \subsubsection{Sub-case: real parameters} 
 It is helpful to begin with
 the sub-case where all the parameters $(\omega,\theta,c$) are real, with $\omega \ge 1$,
 $c \ge 0$ and $\theta$ of any sign. In that case, from (\ref{eq:hatGfirstsol}),
 
 \[ a(s) = r(s) + \psi  = \Smallfrac{1}{2} (1 - \omega) +  \Smallfrac{1}{2} \sqrt{(1-\omega)^2 + 4 s} +
       \psi, \]
  where $\psi = \omega \gamma/R \ge 0$. Notice that $\psi$ is real, non-negative, and 
  \emph{independent of $s$}.
  To find $a(s) = -n$, we must have $r(s) = \Smallfrac{1}{2} (1 - \omega) +  \Smallfrac{1}{2} \sqrt{(1-\omega)^2 + 4 s}$ both real and non-positive. For $r(s)$ to be real,
  we must have $s$ real and lying on the real $s$-axis with $s \ge s_c$. Since $\omega \ge 1$,
  at $s=0$, we  have $r(0) = \Smallfrac{1}{2}[(1-\omega) + (\omega-1)] = 0$. For $s > 0$, we have $r(s) > 0$.
  Hence, $r(s)$ is both real and negative only when $s \in [s_c,0]$. For emphasis: 
  
  \begin{itemize} 
  	\item \emph{under
  real parameters, the only possible poles $($the discrete spectra$)$ lie on the negative real $s$-axis in-between the 
  branch cut at $s = s_c = -(\omega-1)^2/4$ and the origin}. 
  \end{itemize}

The reader should picture a graph of $a(s)$ as $s$ increases from $s_c$ to $0$. Suppose
$\psi>0$, so that $a(0)= \psi > 0$. Suppose $\omega$ is sufficiently large so that
$a(s_c) = (1-\omega)/2 + \psi < 0$. Finally, note that, for $s \in (s_c,0)$,

\[  a'(s) = \frac{1}{\sqrt{(1-\omega)^2 + 4 s}} > 0. \] 
Thus, $a(s)$ is monotone increasing over $s \in [s_c,0]$ and, under the assumptions, crosses the
$s$-axis exactly once within the interval. That crossing point is the location of the pole
$s_0$ where $a(s_0) = 0$. 

Under these same assumptions, is there a pole $s_1$ -- i.e., a value
of $s$ where $a(s_1) = -1$? For this to be true, $\omega$ would have to be even larger, now
sufficiently large so that $a(s_c) =  (1-\omega)/2 + \psi + 1 < 0$. If this last relation holds,
then by the same argument, there is a unique point $s_1 \in [s_c,0]$ such that  $a(s_1) = -1$.
It may be helpful to think of the relation $a(s_c) =  (1-\omega)/2 + \psi + 1 < 0$ as simply
a version of our previous relation (for $s_0$) except we have $\psi \ra \psi + 1$. 
Moreover, by picturing how the graph of $a(s)$ shifts vertically as $\psi \ra \psi + 1$, one sees
that, when the putative $s_1$ exists, then \emph{so does $s_0$ and $s_1 < s_0$}. 

The case of general $n$ for $n$ any positive integer should be clear. To have $a(s_n) = -n$,
$\omega$ must be sufficiently large so that  $a(s_c) =  (1-\omega)/2 + \psi + n < 0$.  
When that last relation holds, there is a unique point $s_n \in [s_c,0]$ such that  $a(s_1) = -n$.
Moreover, when the putative $s_n$ exists, then so do all the $s_m$ for $m = 0, 1, \cdots, n-1$.
In addition, these real poles are ordered in the permissible interval $[s_c,0]$ such that

\be  s_c \le s_n < s_{n-1} < \cdots s_0 \le 0. \label{eq:realpoleordering} \eb
What is the condition for at least one pole? As we have seen, for $s_0$ to exist, we must
have $(1-\omega)/2 + \psi < 0$; i.e., 

\be \mbox{for a discrete spectrum to exist, one needs} \quad (\omega -1)/2 > \psi \ge 0.
     \label{eq:singlepolecondition} \eb
     As promised, (\ref{eq:singlepolecondition}) shows there is no discrete
     spectrum for Case 2 $(0 \le \omega < 1$) -- at least for the real parameter sub-case.

What is  $n_{max}$, the largest value of $n$ in (\ref{eq:realpoleordering})? It is the largest integer such that $(1-\omega)/2 + \psi + n < 0$. Let $\lfloor x \rfloor$ denote the
largest integer contained in $x$ under the assumption that $x \ge 0$.  Then we have found that

\be \mbox{provided (\ref{eq:singlepolecondition}) holds},
	     \quad n_{max} = \lfloor (\omega-1)/2 - \psi \rfloor \ge 0.  \label{eq:nmax} \eb
	  Finally, provided (\ref{eq:singlepolecondition}) holds and $n \le n_{max}$, what
	  is $s_n$? The relation $a(s_n) = -n$ is easily solved for $s_n$ to yield
\be s_n = (n + \psi)^2 + (1 - \omega)(n+\psi) 
	  \quad (n = 0,1,\cdots,n_{max}).  \label{eq:svalsNEW} \eb
	  
	  \Pbreak
	  Summarizing at this point we have now developed (\ref{eq:ResThrm2}) to

	\be G(\tau,y,y')  =   G_c(\tau,y,y') +  1_{\{\frac{\omega-1}{2} > \psi\}}
	\sum_{n=0}^{n_{max}} \e^{s_n \tau} 
	\mbox{Res}[\hat{G}(s_n,y,y'] \label{eq:ResThrm3}, \eb
	where $s_n$ is found in (\ref{eq:svalsNEW}) and $n_{max}$ is found in (\ref{eq:nmax}).   
	It remains to calculate the required residues at these poles.

\subsubsection{Residues}
From (\ref{eq:hatGfirstsol}), since $a(s) = r(s) + \psi$, we have $r_n \equiv r(s_n)= -(n+\psi)$.
Since $b(s) = \omega + 2 r(s)$, we have $b_n \equiv b(s_n) = \omega + 2 r_n$.  Recalling
that $M(a,b,z)/\Gamma(b)$ is regular everywhere, we have
\[ \lim_{s \ra s_n} \frac{1}{\Gamma(b_s)} M(a_s,b_s,\zeta) = \frac{1}{\Gamma(b_n)} M(-n,b_n,\zeta), \]
where $\zeta$ denotes either $R y$ or $R y'$. Also, from the $U$ representation (\ref{eq:Udef}),
\[ \lim_{s \ra s_n} U(a_s,b_s,\zeta') = \frac{\Gamma(1-b_n)}{\Gamma(1-n-b_n)} \, M(-n,b_n,\zeta'), \]
since the pole in $\Gamma(a(s_n))$ suppresses the second factor of $M$. Here $\zeta'$ denotes
$R y'$ when $\zeta = R y$ and vice-versa.

Recall that, as $z \ra -n$,
\[  \Gamma(z) \approx \frac{(-1)^n}{n!} \frac{1}{z + n}, \quad (n = 0, 1, 2, \cdots).
      \]   
      Thus, as $s \ra s_n$
      \[ \Gamma(a(s)) \approx \frac{(-1)^n}{n!}\frac{1}{a'(s_n)(s - s_n)} 
      = \frac{(-1)^n}{n!}\frac{(\omega-1-2n -2 \psi)}{(s - s_n)}.   \]
 Putting it all together we have, as $s \ra s_n$,
  \begin{empheq}{align*}    
    \hat{G}(s,y,y') &\approx   \frac{(-1)^n}{n!}\frac{(\omega-1-2n -2 \psi)}{(s - s_n)}
    \frac{(b_n)_n}{\Gamma(b_n)} R^{\omega-1+ 2 r_n} y^{r_n} (y')^{r_n + \omega-2}
      \e^{-\gamma y} \e^{-(\gamma + \theta) y'}  \\
         &\times	M(-n,b_n, R y) \, M(-n,b_n, R y'), \\
         \end{empheq}
  where we have written $\Gamma(1-b_n)/\Gamma(1-n-b_n) = (-1)^n (b_n)_n$ using the Pochhammer symbol. 
  Factoring out the stationary density, we have found that
  
   \begin{empheq}{align*}    
  \mbox{Res}[\hat{G}(s,y,y')] &= R \, (R y')^{\omega-2} \e^{-\theta y'} 
   	\frac{(\omega-1-2n -2 \psi) \, (b_n)_n}{n! \, \Gamma(b_n)}  \\
  	&\times	 (R^2 y \, y')^{r_n} 
  	\e^{-\gamma (y+y')} M(-n,b_n, R y) \, M(-n,b_n, R y'). \\
  \end{empheq}  
Summarizing, with real parameters, we have fleshed out (\ref{eq:ResThrm2}) to read
\begin{empheq}[box=\fbox]{align}
	&G(\tau,y,y') = G_c(\tau,y,y') \nonumber \\
	\quad &+	R \, (R y')^{\omega-2} \e^{-\theta y'} \e^{-\gamma (y+y')} 
	\times 1_{\{\frac{\omega-1}{2} > \psi\}}
	\sum_{n=0}^{ \lfloor (\omega-1)/2 - \psi \rfloor} \phi_{n}(y,y') \, \e^{s_n t}  \label{eq:Gtotalreal} 
\end{empheq}
where
\begin{empheq}{align*}
	\phi_{n}(y,y') &= \frac{(\omega-1-2n-2\psi) \, (b_n)_n}{n! \, \Gamma(b_n) } 
	\,  (R^2 y \, y')^{r_n} M(-n,b_n, R y) \, M(-n,b_n, R y'), 
\end{empheq}
and using the other notations above. The expression (\ref{eq:Gtotalreal}) completes
the proof of Theorem 1 for the case of real parameters.

\subsubsection{General case: complex parameters} 
As we saw in Sec. \ref{sec:reduction}, complex parameters arise
under the reduction of the bivariate XGBM model to (\ref{eq:Greenfunc}).
For example, recall what happens when trying to calculate option values under the
risk-neutral (Q) model. Then, one has 
the mapping of equation (\ref{eq:qparms}).
That is, $(\theta,c) \ra (\theta_z^+,c_z^-)$ where $z$ is in the preferred Fourier
inversion strip ($0 < \Im z < 1$) and
\be \theta_z^+ = \Smallfrac{2}{\xi^2}  (\theta + \i z \rho \xi), \quad c_z^- =  \Smallfrac{1}{\xi^2}   
\left( z^2 - \i z \right). \label{eq:qparms2} \eb
With these changes, $\hat{G}$ in (\ref{eq:hatGfirstsol}) still
`works'. Since $\omega$ remains real under these complexifications, the
spectral solution components $G_c(\tau,y,y')$ remains the same:
the location of both the branch point at $s = s_c = -(1-\omega)^2/4$ and the associated branch
cut is unchanged.\footnote{More carefully, by `the same', I mean 
	the \emph{formula} for  $G_c(\tau,y,y')$ continues to be (\ref{eq:GspectralCase2}),
	already derived allowing complex values for $(R,\gamma,\psi)$.} 
	 What does change is that the formerly non-negative parameter $\psi$ should now
be considered complex; it remains $s$-independent. Now at  fixed $z$:
\[ \psi \ra \psi_z = \frac{\omega \gamma_z}{R_z} = \frac{\omega}{2}  
      \left( 1 - \frac{\theta_z^+}{\sqrt{(\theta_z^+)^2 + 4 c_z^-}} \right) \]  
  
  With complex $\psi_z$, let's write again the pole criteria (\ref{eq:polecriteria}), which reads     
   \[ \mbox{\underline{pole criteria}:} \,\, s = s_n, \,\, \mbox{such that} \]
 \be  \sqrt{(\omega-1)^2/4 + s} =
    \Smallfrac{1}{2} (\omega-1) -n - \psi_z, \,\, 
   (n=0,1,2, \cdots) \,\,  \mbox{and} \,\, s \in \Omega. \label{eq:ploecriteria2} \eb
For any complex number $\zeta$, then $\sqrt{\zeta}$ has a branch point singularity
at the origin. With the branch cut conventionally placed along the
negative $\zeta$-axis, then in the cut plane $\zeta = R \, \e^{\i \phi}$ with $-\pi < \phi < \pi$.
Thus $\sqrt{\zeta} = \sqrt{R} \, \e^{\i \phi/2}$, which shows $\Re \, \sqrt{\zeta} > 0$ for all $R >0$.
This applies to our situation (\ref{eq:ploecriteria2}) with the identification $\zeta = s - s_c = 
s + (\omega-1)^2/4 $. In other words, since  $\Re \, \sqrt{\zeta} > 0$ in the cut $s$-plane, we
can take the real part of both sides of (\ref{eq:ploecriteria2}) to find that
\begin{empheq}{align}
	 \Smallfrac{1}{2} (\omega-1) - n - \Re \, \psi_z > 0 \,\, \Rightarrow \,\, 
	n <  \Smallfrac{1}{2} (\omega-1) - \Re \, \psi_z \label{eq:ploecriteria3}
	\end{empheq}
For a non-empty discrete spectrum, we need a solution $s \in \Omega$ to (\ref{eq:ploecriteria2}) for $n=0$. That
requires $\Smallfrac{1}{2} (\omega-1) > \Re \, \psi_z$ to be satisfied. The poles themselves
are still given by the same formula (\ref{eq:svalsNEW}), now
\begin{empheq}[box=\fbox]{align}
  s_n &= s_n(z) =  (n + \psi_z)^2 + (1 - \omega)(n+\psi_z) \label{eq:svals2},
\quad (n = 0,1,\cdots,n_{max}), \\
  &\mbox{where} \quad n_{max} = \lfloor \Smallfrac{1}{2} (\omega-1) - \Re \, \psi_z \rfloor. 
        \nonumber
 \end{empheq}       
The same formulas as before still hold for the residues. Thus, we have the
generalization, with complex parameters:
\begin{empheq}[box=\fbox]{align}
	&G(\tau,y,y') = G_c(\tau,y,y') \nonumber \\
	\quad &+	R \, (R y')^{\omega-2} \e^{-\theta y'} \e^{-\gamma (y+y')} 
	\times 1_{\{\frac{\omega-1}{2} > \Re \psi\}}
	\sum_{n=0}^{ \lfloor (\omega-1)/2 - \Re \psi \rfloor} \phi_{n}(y,y') \, \e^{s_n \tau}  \label{eq:Gtotalcomplex} 
\end{empheq}
where
\begin{empheq}{align*}
	\phi_{n}(y,y') &= \frac{(\omega-1-2n-2\psi) \, (b_n)_n}{n! \, \Gamma(b_n) } 
	\,  (R^2 y \, y')^{r_n}	M(-n,b_n, R y) \, M(-n,b_n, R y'). 
\end{empheq}
Comparing (\ref{eq:Gtotalcomplex}) to the previous (\ref{eq:Gtotalreal}), the formulas are
very close -- the only change is $\psi \ra \Re \, \psi$ in the pole condition and sum limit.

 \pbold{Complications induced by the Fourier inversion.} When using the above formulas for the bivariate model, as we
 have discussed, the parameter $\psi_z$ depends upon the
 Fourier parameter $z$. For each fixed $z$, when $\omega -1 > 2 \, \Re \psi_z$,
  the number of discrete eigenvalues $N_e(z)$ is
 given from the above by $N_e(z) = 1 +  \lfloor (\omega-1)/2 - \Re \psi_z \rfloor$. 
 Thus, when integrating along a contour in the complex $z$-plane to implement
 (\ref{eq:putval}), you will often find $N_e(z)$ changing by unit steps
 along the integration contour. 
 
\begin{figure}[h] 
 	   	\caption{\textbf{Discrete Spectrum Size along a Fourier Integration Contour}}
 	   	\vspace{10pt}
 	   	\includegraphics[width=0.9\textwidth]{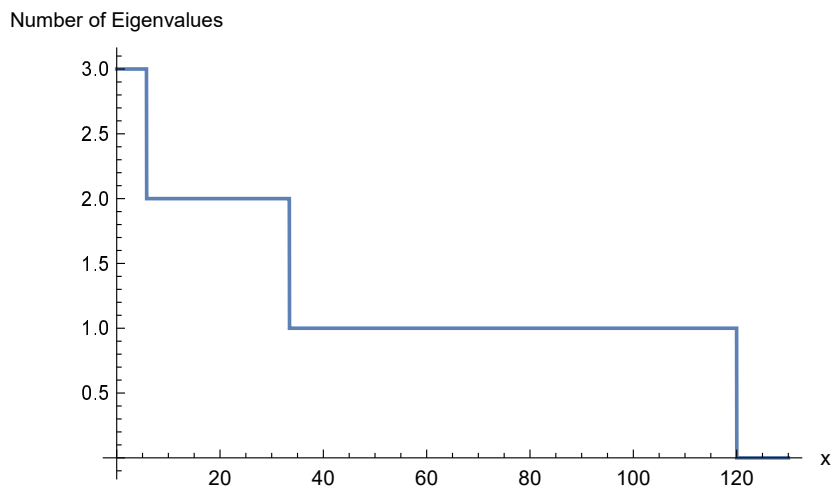}
 	   	\label{fig:NumEigenPlot}
 	   \end{figure}

 An example should make this clear. Take the risk-neutral XGBM option model with $\omega = 3$, $\theta = 4$,
 $\rho = -0.8$, and $\xi=1$. To find option values, we perform a Fourier inversion by integrating along
  $z = x + \i/2$ for $x \in [0,\infty)$. Fig. \ref{fig:NumEigenPlot} plots the
  number of eigenvalues (number of elements in the discrete spectrum) as $x$ varies
  along this contour. As you can see, at the beginning of the contour ($x=0$),
  there are 3 eigenvalues in the spectrum, but they each subsequently drop out as $x$ increases,
  at \emph{transition points} $x \approx (5.80,33.42,120.02)$. Thus, at large $x$, indeed for
  all $x > 120.02$, the discrete spectrum is empty. 
  
\begin{figure}[hp] 
 		   	\caption{\bf{Eigenvalue Trajectories $s_n(z)$ in the Complex $s$-plane vs. $z$}}
 		   	\vspace{10pt}
 		   	\begin{center}
 		   		\includegraphics[width=0.8\textwidth]{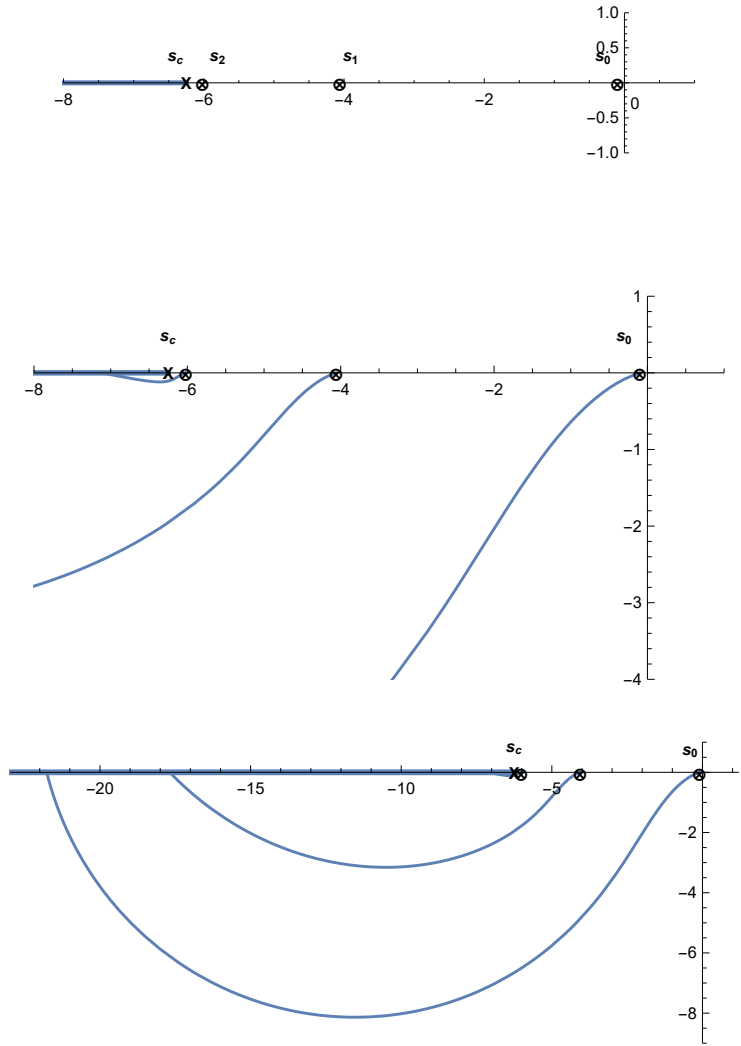}
 		   		\newline\newline
 		   	\end{center}		
 		   	\label{fig:SpectrumTrajectories}
 		   	{\bf{Notes.}} Parameters: $\omega=3$, $\theta = 4$, $\rho = -0.8$, and $\xi = 1$. Fourier integration is along the contour $z = \i/2 + x$, where at most 3 eigenvalues $s_n(x)$, $(n=0,1,2)$ contribute. Top chart shows the $s_n(0)$. Bottom chart shows parametric plots of $s_n(x)$ as $x$ increases along the contour. Middle chart shows detail to make the $s_2(x)$ trajectory clearer. The trajectories touch the branch cut and the corresponding eigenvalues disappear from the solution at $x_n \approx \{5.796,33.424,120.017\}$ for $\{s_2,s_1,s_0\}$ respectively. 
 		   \end{figure}

\begin{figure}[hp] 
 		   	\caption{\bf{Continuous, Discrete, and Full Fundamental Transform:} \newline $\Re \, 
 		   		H(z = i/2 + x)$ vs. $x$}
 		   	\vspace{10pt}
 		   	\begin{center}
 		   		\includegraphics[width=0.6\textwidth]{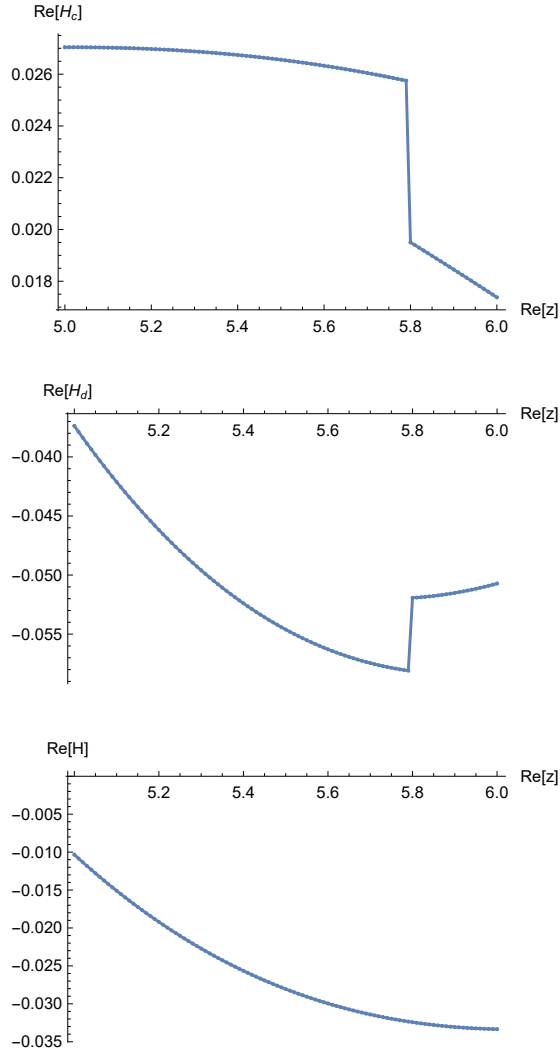}
 		   		\newline\newline
 		   	\end{center}		
 		   	\label{fig:TransitionPt}
 		   	{\bf{Notes.}} Model parameters as in Fig.\ref{fig:SpectrumTrajectories}. The trajectory for the eigenvalue $s_2(z)$ along the Fourier integration contour $z = \i/2 + x$ touches the branch cut at $\Re z = x_2 \approx 5.796$. At that point the fundamental transform $H(z) = H_c(z) + H_d(z)$ is regular (smooth), while the continuous and discrete components, $H_c(z)$ and $H_d(z)$ respectively, have offsetting jumps. This  behavior occurs whenever a discrete eigenvalue $s_n(z)$ touches the $s$-plane branch cut as it leaves the solution.	 
 		   \end{figure}

  What is going on that causes the drop-outs?  
 If you look at trajectory plots
 of $s_n(z)$ in the complex $s$-plane (as $z$ varies), you will find that the
 drop-outs occur precisely where the $s_n(z)$ \emph{touch the branch cut} (Fig. \ref{fig:SpectrumTrajectories}).

 Note that the branch-cut itself is \emph{not} in the region $\Omega$. Thus, the qualification $s \in \Omega$ in (\ref{eq:ploecriteria2}) is important and is enforced by the behavior we have just described.
 Eigenvalues reaching the branch cut is a new feature of the complex parameter case. With real
 parameters, $s_n(z)$ can approach the cut -- but only from the real axis. So, if it
 reached the cut, it would reach it at the branch point end-point only, where the formulas tend to be 
 smoother. 
 
 A complication associated with this behavior is that the two components of the 
 spectral expansions above (discrete and continuous) can show discontinuities
 at these transition points -- even though their sum is smooth.   
 While $G(\tau,y,y';z)$ itself will be smooth as $z$ passes
 through a transition value, there may
 be off-setting discontinuities in the (continuous and discrete) components $G_c$ and $G_d$ or associated functions. Examples of associated functions to $(G,G_c,G_d)$ are the marginals
 $(H,H_c,H_d)$ in which the terminal volatility $y'$ is integrated out. See
 	Fig. \ref{fig:TransitionPt} for examples of how the branch
 	cut touches of Fig. \ref{fig:SpectrumTrajectories} manifest themselves in discontinuities
 	of $H_c$ and $H_d$, while $H$ remains continuous in $z$. This behavior needs to be carefully handled in
 	any coding implementation: we discussed our particular method in Sec. \ref{sec:numerics}.

 \newpage
 \section{Appendix B -- The SABR model limit}
 
 If one drops the drifts from (\ref{eq:XGBM1}), the system becomes
 
 \bea  \left \{ 
 \begin{array}{ll}
 	d S_t =  \sigma_t S_t \, dB_t,  \quad      &  S_t \in \RBB_+, \\
 	d \sigma_t = \xi \, \sigma_t \, dW_t,  \quad & \sigma_t \in \RBB_+, \\
 	dB_t \, dW_t = \rho \, dt,          
 \end{array}  \right.  \label{eq:SABR}  
 \aeb
 which is the lognormal SABR model. Equivalently, (\ref{eq:XGBM2}) becomes  
 \bea   \left \{ 
 \begin{array}{ll}
 	d X_t =  -\frac{1}{2} Y^2_t  \, dt + Y_t \, dB_t,  \quad    &    X_t \in \RBB, \\
 	d Y_t =  \xi \, Y_t \, dW_t,  \quad & Y_t \in \RBB_+, \\
 	dB_t \, dW_t = \rho \, dt.          
 \end{array}  \right.  \label{eq:SABR2}  
 \aeb
 Now, (\ref{eq:solnfinal}) holds with $\omega=0$,
 $\theta_z^- = -2 \i z \rho/\xi$, and $c_z^+ = (z^2 + \i z)/\xi^2$. Tracing the implications in Theorem \ref{thm:Green}, there are considerable simplifications. Using the notations there, since
 $\omega = \psi = 0$, there is no discrete spectrum contribution. In the continuous spectra integrand, the key simplification
 is the relation (see \cite{as:1972}):
 
 \[ U(\Smallfrac{1}{2} + \i \nu, 1 + 2 \i \nu, z) = \frac{\e^z z^{-\i \nu} }{\sqrt{\pi}}\,  
       K_{\i \nu} \left( \frac{z}{2} \right), \]
 	where $K_{\nu}(z)$ is a modified Bessel function. One can also show from  \cite{as:1972} that
 	\[ \left| \frac{\Gamma(\frac{1}{2} + \i \nu)}{\Gamma(2 \i \nu)} \right|^2 = 4 \nu \sinh \pi \nu. \] 
 	Putting it all together, (\ref{eq:solnfinal}) reduces to
 	
 	\begin{empheq}{align}
 		&\, p_{SABR}(t,x',y'|x,y) = 
 		\frac{1}{\pi^3} \, \frac{y^{1/2}}{(y')^{3/2}} \,\, \e^{-\frac{1}{8} \xi^2 t}  \,\, \times 
 		\label{eq:pSABRalt} \\
 		&  \int_{-\infty}^{\infty} \int_0^{\infty} \e^{-\frac{1}{2} \nu^2 \xi^2 t}
 		\, \nu  \, (\sinh \pi \nu) \,
 		K_{\i \nu}(\chi_z \, y) \, K_{\i \nu}(\chi_z \, y')  
 		\,  \, \e^{-\i z[(x-x') - (\rho/\xi) (y-y')]}  \, d \nu  \, d z, \nonumber  \\
 		&\, \mbox{using} \,\, \chi_z = [(1-\rho^2) z^2 - \i z]^{1/2}.  \nonumber
 	\end{empheq}

 \vspace{-40pt}
 \Pbreak
Eqn (\ref{eq:pSABRalt}) agrees with a representation for the SABR transition density previously found in \cite{lewis:2016} (Ch. 8, eqn (8.81)). Note that in (\ref{eq:pSABRalt}) I have flipped the sign of dummy integration variable $z$ relative to (\ref{eq:solnfinal}) to make the comparison with  \cite{lewis:2016}.

 	\newpage
 	\section{Appendix C -- The martingale defect} 
 	
 	The XGBM model under the risk-neutral (Q-measure) evolution
 	is given at (\ref{eq:XGBM3}). Using $(\omega_Q,\theta,\xi,\rho)$ from there, define $\beta = 2 - 2 \, \omega_Q/\xi^2$,
 	 and $\gamma = 2 [(\rho/\xi)-(\theta/\xi^2)]$. We also need $a_{\mu}=(\beta-1)/2+ \i \mu$, $c_{\mu}=1+2 \i \mu$, 
 	 $c_{m}=\beta-2-2m$, and $d_{m}=c_{m}-1$. 
 	 Then, one can show that the martingale defect under that model is given by 
 	 \be  \Esub{S_0,\sigma_0}{\frac{S_t}{S_0}}
            =  
 	 \left\{\begin{array}{ll}
 	 	1     \quad & \gamma \le 0 \quad (\mbox{i.e.,} \, -1 \le \rho \le \theta/\xi),                     \\
 	 	 1 - A \left(\sigma_0,t \right), \quad &  \gamma > 0 \quad (\mbox{i.e.,} 
 	 	 \,\,\,  \theta/\xi < \rho \le 1).  \\
 		\end{array}\right. \label{eq:mdefect1} \eb
  	Here $A(x,t)$ is an absorption probability, given by:
 
 		\be A(x,t)=A_{\infty}(x) + A_{d}(x,t) + A_{c}(x,t),  \eb
 		\be \mbox{where} \quad A_{\infty}(x)=\left\{\begin{array}{ll}
 			1-\frac{\Gamma(\beta-1,\gamma x)}{\Gamma(\beta-1)}, \quad  & \beta > 1, \label{eq:ae221} \\
 			1, \quad & \beta < 1, \\
 		\end{array}\right.  \eb
 		\bea A_{d}(x,t) &=& 1_{\{\beta>3\}} \, \e^{-\gamma x} \sum_{m=0}^{ \lfloor(\beta-3)/2 \rfloor} 
 		\frac{(-1)^{m+1}}{m!} \frac{(\gamma x)^{\beta-2-m}}{\Gamma(d_m)} \nonumber \\
 		&\times&   
 		\frac{\exp\{-\frac{1}{2}[\beta-2+m(\beta-3-m)] \xi^2 t\}}{[\beta-2+m(\beta-3-m)]}
 		M \left( -m,c_{m},\gamma \, x \right), \quad\quad\quad \aeb
 		\be \mbox{and} \quad A_{c}(x,t)= -\int_{0}^{\infty} \overline{g}(\mu) \, \eta 
 		 \left( \gamma \, x, \mu \right)
 		\exp \left\{ -\Smallfrac{1}{2} \left[ \mu^2 + \frac{(\beta-1)^2}{4} \right] \xi^2 t \right\} d \mu, \eb
 		\be  \mbox{using} \quad \bar{g}(\mu)=\frac{1}{2 \pi}
 		 \left| \frac{\Gamma(c_{\mu}-a_{\mu})}{\Gamma(2 \i \mu)} \right|^{2}
 		\frac{1}{\mu^{2}+\frac{(\beta-1)^{2}}{4}},  \nonumber \eb
 		\be \mbox{and} \quad \eta(y,\mu) = 
 		\e^{-y} \, y^{a_{\mu}} \, U(c_{\mu}-a_{\mu},c_{\mu},y). \nonumber \eb
 		\label{prop:ae3}
 		
 		\Pbreak
 		Proof: Follows easily from the absorption probability found in Sec 13.2.9 of
 		   Ch. 13 of \cite{lewis:2016}. 
 		   
 		   \pbold{Remarks.} As discussed in \cite{lewis:2000a}, an alternative route to
 		   the martingale defect is the relation $\Esub{S_0,\sigma_0}{\frac{S_T}{S_0}} = H(T,\sigma_0;z=i)$.
 		   Showing the equivalence of this last relation to (\ref{eq:mdefect1}) is
 		   a good exercise for the interested reader when $T \ra \infty$. A direct check for general 
 		   $T$ has eluded us. 
 		   
 		   \newpage

 		   \newpage
 		   
 		   \newpage
 		   
 	   \newpage
 	   
\end{document}